\documentclass[11pt]{article}
\usepackage[super,sort&compress]{natbib}
\usepackage{fourier}
\usepackage[utf8]{inputenc}
\usepackage{booktabs}
\usepackage{graphicx} 
\usepackage{amsmath, amsfonts, amssymb}
\usepackage{float}
\restylefloat{table}
\usepackage{pdflscape}
\usepackage[a4paper]{geometry}
\allowdisplaybreaks
\usepackage[bookmarks]{hyperref}
\usepackage{mathtools}
\usepackage[table,xcdraw]{xcolor}
\usepackage{placeins}
\usepackage{anyfontsize}
\usepackage{multicol, multirow}
\usepackage{tablefootnote}
\usepackage{threeparttable}
\usepackage{tikz}
\hypersetup{
    colorlinks,
    linkcolor={red!20!black},
    citecolor={blue!50!black},
    urlcolor={blue!80!black}
}
\usepackage{titlesec}

\AtBeginDocument{%
  \RenewCommandCopy{\leq}{\leqslant}%
  }
  \newcommand{\logit}{\mathrm{logit}}

\titleformat{\paragraph}
  {\normalfont\normalsize\bfseries}{\theparagraph}{1em}{}
\titlespacing*{\paragraph}
{0pt}{3.25ex plus 1ex minus .2ex}{1.5ex plus .2ex}

\titleformat{\paragraph}
{\normalfont\normalsize\bfseries}{\theparagraph}{1em}{}

\title{When Bayes goes bad:  Weakly-regularized covariate adjustment leads to a biased estimate of prevalence}
\author{Swen Kuh\footnote{School of Mathematical Science, Adelaide University -- Tirkangkaku, Australia} \and Lauren Kennedy\footnotemark[1] \and Qixuan Chen\footnote{Department of Biostatistics, Columbia University, New York} \and Andrew Gelman\footnote{Departments of Statistics and Political Science, Columbia University, New York}} 

\date{12 Mar 2026}

\begin{document}

\maketitle

\begin{abstract}
When estimating population prevalence from a non-random sample, it is important to adjust for differences between sample and population. However, adjustment for multiple factors requires analysis that can be difficult to understand and validate. In this manuscript, we explore an unexpected downward trend of estimates when covariates are added sequentially to a Bayesian hierarchical model for the estimation of the prevalence of SARS-CoV-2 specific antibodies in an Australian city in late 2020.

We compare our data analysis to results from a simulation study to understand four potential contributors to this effect: (i) correction for differences between sample and population, (ii) rare-events bias in logistic regression, (iii) inclusion of the uncertainty of test sensitivity and specificity in a multilevel model, and (iv) increasing model dimensionality. We find that weak prior distributions on the logistic regression coefficients lead to a systematic increase in the amount of partial pooling across adjustment cells---the prior becomes stronger as model dimensionality increases---which in turn feeds through to the estimated assay specificity, which then feeds back to the model and results in lowering the estimated prevalence.

Our paper contributes three elements: (i) immediate and longer-term recommendations for using these types of models, (ii) simulation studies to explore the impact of the contributors to this effect, and (iii) a worked example of investigation of unexpected results in a model with multiple adjustment factors.
\end{abstract}

\graphicspath{ {./figures/} }

\maketitle

\section{Introduction}
Bayesian inference allows us to build models that account for multiple sources of uncertainty, and in complex problems this can produce surprising or counterintuitive estimates. How can we diagnose the difference between  valid yet surprising inferences, and artifacts caused by inappropriate models? In this work, we explore a problem that arose when aiming to estimate the prevalence of SARS-CoV-2 (COVID-19) antibodies in metropolitan Melbourne, Australia in Nov--Dec 2020 based on a serosurvey of blood donors.

In 2020, Australia attempted an elimination strategy for COVID-19 through strict border controls both domestically and internationally. Although most states achieved and maintained zero or very few COVID-19 cases in this period, the city of Melbourne experienced several outbreaks and was attempting to control them through movement restrictions and other containment measures. Case counts were derived from free and accessible tests used if symptoms were present or if notified as a ``close contact'' of an infected individual. However, there remained uncertainty regarding the underlying incidence of COVID-19 especially because infection could be asymptomatic. In an attempt to quantify this, a range of measures were used, including waste treatment screening,\citep{black2021} but here we focus on an attempt to monitor the prevalence of SARS-CoV-2 infection based on residual blood specimens from blood donors (seroprevalence). 

The analysis by Machalek et al.\citep{machalek2022} aimed to estimate seroprevalence for two populations: (i) Melbourne metropolitan blood donors and (ii) Melbourne metropolitan residential population from a sample of residual blood specimens. This work focuses on the results and estimation issues specific to the residential population. In order to obtain the sample, a stratified sampling design was used based on three postcode groups defined by low (less than 3 cases per 1000 population; sample size $n = 1,600$), medium (3--7 cases per 1000 population; $n = 1,600$), and high (more than 7 cases per 1000 population; $n = 1,599$) incidence of COVID-19 based on case notification data. The resulting data contained 4,799 samples from blood donors aged 20--69 years that were collected in Melbourne during Nov--Dec 2020.  

Test sensitivity and specificity were estimated using independent Wantai SARS-CoV-2 total antibody assay calibration data by assessing $102$ stored specimens that were confirmed RT-PCR positive and collected more than 14 days post-symptom onset. Of these, 97 gave a positive result, yielding an estimated test sensitivity of $97/102 = 95.1\%$. Test specificity was assessed by using 800 (pre-pandemic) blood donor specimens from May 2019. Three positives were found, giving an estimated test specificity of $797/800 = 99.6\%$.\citep{vette2022} 

In the analysis for this project, the statisticians on the team noticed that the overall estimate of COVID-19 antibody prevalence decreased as they added more adjustment variables to their multilevel model (further described in the following section). While this might be because the estimate was moving closer to the truth as the model grew more complex and accurate, it was surprising (as we detail later) and warranted closer inspection and validation.

Our manuscript proceeds as follows. In the remainder of the introduction, we detail the model used, the motivating example, and our proposed workflow for investigation. We then present a series of simulation studies and investigations that aim to provide evidence for different hypothetical causes. 

\subsection{Multilevel modeling to estimate disease prevalence}

Gelman and Carpenter\citep{gelman2020} proposed a method to simultaneously address imperfect diagnostic test performance and non-representative samples when estimating disease prevalence. They use a multilevel regression and poststratification (MRP) method to address differences between the sample and the population while also incorporating estimation uncertainty in test sensitivity and specificity. We describe their approach here by first describing MRP and then modifications to a typical MRP method to incorporate (a) test sensitivity and specificity, and (b) allowing estimation of sensitivity and specificity. Using one modeling framework for all parts of the problem is important as it allows all sources of uncertainty to be propagated through to the final prevalence estimate. 

We begin by defining notation where medical screening tests provide a proxy for the true presence of disease. In this case we observe $y^*$ (the detection of SARS-CoV-2 antibodies from the diagnostic assay) and not $y$ (the true SARS-CoV-2 antibody status of the patient). We denote the observed test outcome as $y_i^*$, and the expected probability that the $i^{th}$ individual tests positive as $\mbox{Pr}(y_i^* = 1)$ (in contrast to $\pi_i = \mbox{Pr}(y_i = 1)$, the true status). If the sensitivity $\delta = \mbox{Pr}(y_i^* = 1|y_i =1)$ and the specificity $\gamma = \mbox{Pr}(y_i^* = 0|y_i =0)$ are both  high, then we would expect that $\mbox{Pr}(y_i^* = 1)$ is close to $\pi_i$.  

In a general case (where we would observe $y$ directly) MRP follows a two-step process.  First, the relationship between the outcome (in this case, whether SARS-CoV-2 specific antibodies are present in a sample of blood) and participant demographics (for example, variables such as sex, geographic location and age group) are modeled using a multilevel model. Although the aim is population estimation, a  sample may be used for this model under the assumption of exchangeability within each demographic combination. Second, this model is used to make predictions for each demographic combination at a population level. These predictions are then aggregated to infer a population-level or sub-population-level estimate. This method has been widely applied across disciplines, including political science,\citep{lax2009should, buttice2013} health sciences,\citep{zhang2015, downes2018} and survey research.\citep{si2017}

The model used in MRP for a binary outcome is most commonly a multilevel logistic regression. We split the set of demographic predictors into $K$ covariates with more than two levels (e.g., age group), modeled using a varying effect, and $Q$ covariates with exactly two levels (e.g., sex) modeled with a non-varying (fixed/overall) effect.  We denote $y_i$ as the true SARS-CoV-2 antibody status of individual $i$ and let $\pi_i = \mbox{Pr}(y_i = 1)$ be the probability that $i^{th}$ individual has the SARS-CoV-2 antibodies present in their blood. We model the relationship between the two as
\begin{align*}
\widehat{\mbox{Pr}}(y_i = 1) &= \mbox{logit}^{-1}\big(\hat{\beta}_0 + \sum_{q=1}^Q\hat{\beta}_qx_q + \sum_{k=1}^K\hat{\alpha}^{(k)}_{l[i]}\big)  \label{eq:basic_mrp} \\
\hat{\alpha}^{(k)}_{l} &\sim \mbox{normal}(0,\hat{\sigma}_k).
\end{align*}
where $\beta_0$ represents an intercept term, $\beta_q$ represents a non-varying effect for the $q^{th}$ variable. The term $\alpha^{(k)}_{l[i]}$ represents the value of $\alpha$ for the given the individual $i$ is in level $l$ of the $k^{th}$ varying effect. We denote the total number of levels in the $k^{th}$ varying effect as $L^{(k)}$. (When the $k^{th}$ variable is binary, the variance, $\sigma_k$ cannot be estimated. Instead a single parameter is included as offset an offset from the intercept term.)

The population can be described by creating a data frame where each one row represents one combination of demographics (e.g., women aged 18--25 with high school education). The total number of rows possible is $J = 2Q\prod_{k=1}^KL^{(k)}$.  Through a population census or other high-quality survey, we can assume $N_j$, the number of people in the population in each demographic combination, is known. Using the assumed exchangeability within the $j^{th}$ demographic category, we know that we estimate the same $\widehat{\mbox{Pr}}(y_i = 1)$ for all $i \in j$. For shorthand, we denote this as $\widehat{\mbox{Pr}}(y_j = 1)$. From this, a population estimate can  be created:
\begin{align}
    \widehat{\mbox{Pr}}(y = 1) &= \frac{\sum_{j=1}^J N_j \widehat{\mbox{Pr}}(y_j = 1)}{\sum_{j=1}^J N_j }.
\end{align}
To estimate an unobserved $\pi_i$ from the observed $y^*$, Gelman and Carpenter\citep{gelman2020} exploit the flexibility of a Bayesian approach to model the true population disease prevalence given demographic variables and then constructing the observed values from this parameter and the test sensitivity and specificity. The form of the model builds on equation (\ref{eq:basic_mrp}) with the additional estimation of
\begin{equation}
\widehat{\mbox{Pr}}(y^*_{i}=1) = (1-\gamma)(1-\pi_i) + \delta\pi_i. \label{eq:merror}
\end{equation} 
Inclusion of sensitivity and specificity when estimating prevalence is well established, especially in the veterinary science literature \citep{brenner1997, branscum2004}, but Gelman and Carpenter \cite{gelman2020} are, to our knowledge, the first to propose propagating the uncertainty of estimating sensitivity and specificity into a multilevel model to adjust for non-random sampling. This method had not been used extensively before deployment by Machalek et al.\citep{machalek2022}, our motivating example. 

Gelman and Carpenter discuss two ways to incorporate prior information regarding test sensitivity and specificity. The first is a meta-analysis approach:  suppose $V$ different studies are available that report sensitivity and specificity for a particular test in a particular sample with size $M_v$. Then we can model the sensitivity and specificity as 
\begin{align*}
    \text{logit}(\delta_v) &\sim \text{normal}(\mu_\delta, \sigma_\delta), \\    
    \text{logit}(\gamma_v) &\sim \text{normal}(\mu_\gamma, \sigma_\gamma),
\end{align*}
respectively. $\mu_\delta$ and $\mu_\gamma$ represents average over studies of sensitivity and specificity, and $\sigma_\delta$ and $\sigma_\gamma$ represent the variance across studies.  

However, in our motivating example, only one dataset is available from which to estimate sensitivity and specificity. In order to specify a prior for the sensitivity and specificity, Machalek et al.\citep{machalek2022} use the approach from Vette et al.\citep{vette2022} Partition an independent calibration study of size $m$ (where true COVID-19 antibody prevalence is known) into cases with the COVID-19 antibody $m_\delta$ and cases without the COVID-19 antibody $m_\gamma$. For the cases with the COVID-19 antibody, let the true positives (TP) be the cases that are correctly identified by the screening test, and false negatives (FN) be the rest. For the cases without the disease, let true negatives (TN) be the cases that the screening test correctly identifies as disease-free, and false positive (FP) be the rest. Then we set the prior distribution for sensitivity and specificity, respectively, as:
\begin{align*}
 \delta &\sim \mathrm{beta(TP, FN)}, \\
 \gamma &\sim \mathrm{beta(TN, FP)}.
\end{align*} 
We use this prior throughout this manuscript.  

\subsection{Motivating example: Applying multilevel modeling to estimate COVID-19 antibody prevalence}

The original analyses by Machalek et al.\citep{machalek2022} specified a model with all demographic main effects included as covariates but no interactions (equivalent to model 5 in this manuscript). However, in exploring the model robustness, it was observed that when covariates were added successively, there was a downward trend of prevalence estimates away from the crude sample proportion toward zero, as in the navy (dark shade) of Figure \ref{fig:ori_label}. The covariates used and the associated number of covariate levels are as in Table \ref{tab:cov_table}. We found it sufficient to explore models 0 through 5 in this manuscript (excluding models with many interactions). 

\begin{table}
    \centering
   \begin{tabular}{ccc}
Model & Covariate added & Number of levels  \\\hline
0     & None (intercept-only)  & 0  \\ 
1     & Sampling strata  & 3    \\ 
2     & Postcode      & 102 \\ 
3     & Socio-economic Index For Areas (SEIFA) & 5  \\ 
4     & Sex    & 2   \\
5     & Age group   & 5  \\ 
6     & SEIFA $\times$ sex, SEIFA $\times$ age group, sex $\times$ age group  & 10, 25, 10  \\                 
7     & Strata $\times$ SEIFA, strata $\times$ sex, strata $\times$ age group   & 15, 6, 15 
\end{tabular}
\caption{\em Sequential addition of covariates in the models, and the corresponding number of levels for each covariate. Model 0 includes only the intercept, while subsequent models add various covariates. Model 0 has no covariates, Model 1 includes sampling strata as a covariate, Model 2 includes both sampling strata and postcode as covariates, and so on. All effects were estimated using varying $x$-intercept terms to model the group-level effects, except for sex (a binary variable), which was modeled as a non-varying term and SEIFA, which was modeled as both a varying and non-varying term. We focus only on models 0 to 5 in this work.}
    \label{tab:cov_table}
\end{table}

There are at least three elements to the modeling approach used in Machalek et al. The first is the estimation of unknown sensitivity and specificity from a separate dataset, and the propagation of uncertainty from that through to the final estimate. Second, the use of MRP implies that we believe that there is a possibility of (a) heterogeneity in the prevalence of COVID-19 antibodies between demographic subgroups and (b) a difference in demographic proportions in the sample when compared to the population. This means we would expect our MRP population estimate to differ from the sample average. Third, a logistic regression is used to estimate the outcome; yet the outcome is rare in the sample. Of the 4799 observations, only 77 actually tested positive for COVID-19 antibodies, which means that we need to consider the potential for rare-events bias. Lastly, this model combines estimation of multiple components (measurement error, population adjustment, and regression estimation) into one framework, which could interact in unexpected ways.

We propose a hierarchical workflow for the investigation of each of the elements independently and in conjunction to better understand the nature of this artifact.  

\subsection{Diagnostic workflow}

In our workflow, we iterate between the exploration of model robustness with the real-world data example and the use of simulation studies to test specific hypotheses that might result in an artifact. To begin with, we first needed to identify whether the changing estimates are something of concern. The purpose of MRP is to adjust from a  sample to create a population estimate. It is used when the sample is likely to produce a biased estimate of the population. This means that including the choice of adjustment variables impacts the accuracy of our model estimates.\citep{kuh2024using} Thus, for an estimate to change as more parameters are added to the model might be expected behavior. 

\subsubsection*{Correctly applying posterior predictive checking}

Posterior predictive checking is used as one step in the Bayesian workflow to validate the appropriateness of a model for data.\citep{gelman2020bayesian} It is important to target the predictive check to the target estimand. MRP analyses are intended for aggregated estimates, in this case, an overall estimate of prevalence. Say we have $B$ posterior draws for each parameter in the model. Our estimate for the $b^{th}$ posterior predictive value of the disease prevalence in cell $j$ is denoted $\hat{\pi}_j^b$. Using this, we can calculate a posterior predictive sample mean as 
\begin{align}
    \widehat{\pi}_s^b &= \frac{\sum_{j=1}^J n_j \widehat{\pi}^b_j}{\sum_{j=1}^J n_j },
\end{align}
where $n_j$ represents the number of individuals in a particular demographic cell in the sample, rather than the population $N_j$. Adjustment variables are included in the model when they relate to both the probability of inclusion in the sample and the probability of the outcome. This means that posterior predictive sample means from MRP should not differ on average from the observed sample mean. 

\begin{figure}
\centering
\includegraphics[width=0.9\textwidth]{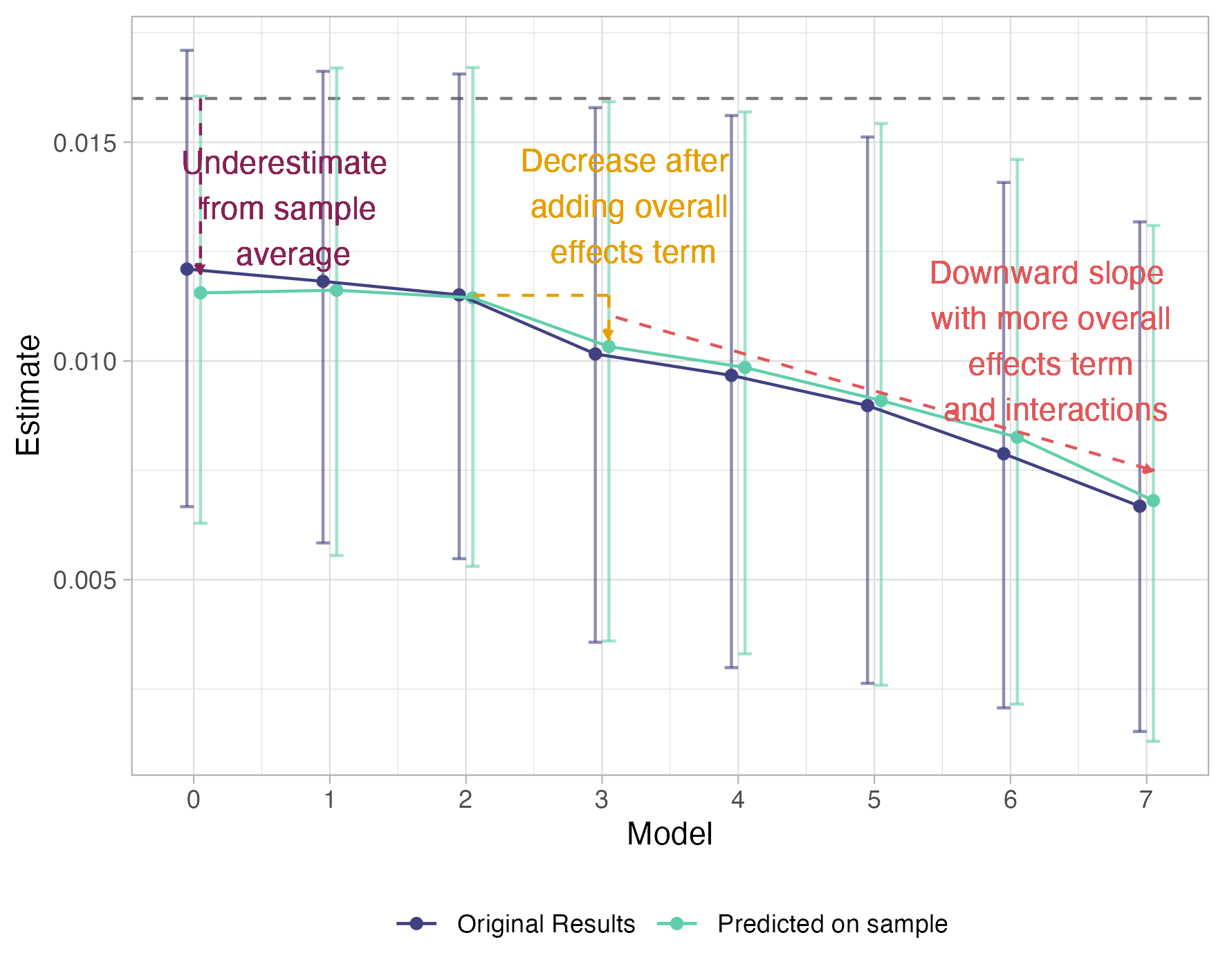}
\vspace{-.2in}
\caption{\em MRP seroprevalence estimate ($y$-axis) for Melbourne metropolitan residential population using models ($x$-axis) with sequentially added covariates as in Table \ref{tab:cov_table}. Color represents the original results creating a prevalence estimate for the population (black) and a prevalence estimate for the sample (light green). The similarity between these two lines suggests that a modeling issue rather than a sample adjustment issue. Uncertainty bars represent 90\% credible intervals. The light gray dotted line indicates the sample mean, while the colored dotted lines and annotations are used to describe the three different challenges.  We focus only on models 0 to 5 in this work.}
\label{fig:ori_label}
\end{figure}

In Figure~\ref{fig:ori_label}, the light green line indicates the posterior predictive sample means (when we predict the sample as if it were the population) from an MRP method. We would expect that if the model is appropriate for the data, we would recover the sample average with little impact from the variables included in the model. However, we see the same downward trend as when we calculate a population estimate. This suggests that the poststratification or population extrapolation does not explain the decrease as we add covariates in the model. Instead, it seems likely that adding covariates is increasing some sort of model misspecification. 

What might cause such an artifact? We annotate Figure~\ref{fig:ori_label} to identify multiple challenges. The purple note on the graph indicates that, even with an intercept-only model, the posterior predictive estimate is an underestimate compared to the observed sample mean, which suggests that even the simplest version of this model is challenged. The orange note flags the decrease of the estimate when adding a variable that is modeled with both varying (random) and non-varying (fixed) effects is added to the model, and the red note points to further decreases as more variables are included in the adjustment.

\subsubsection*{Rare-events bias}

One potential reason for this might be rare-events bias,\citep{leitgob2013,williams2016} which occurs in a logistic regression with small samples in a low-prevalence setting, and results in predicted probability estimates that are, on average, lower than the true value. rare-events bias has been demonstrated for regression coefficients in maximum likelihood estimates for logistic regression.  The problem is also called sparse-data bias\cite{greenland2016sparse} and is closely related to the non-collapsibility of odds ratios in logistic regression.

Nevertheless, our problem is not as simple as small samples and low prevalence. The risk of bias is an interaction between the amount of data and the prevalence of the outcome: logistic regression for small samples with low prevalence can be subject to rare-events bias. Increasing the number of parameters in the model results in fewer data per parameter and can also create this effect, and a rule of thumb has been suggested for 10-20 events per variable (EPV)\citep{peduzzi1995, austin2017}, and up to 50 EPV when variable selection is applied,\citep{steyerberg1999} though further research suggest other factors such as total sample size and proportion of successes also matters.\citep{peduzzi1996, van2016, newsom2021}

In an MRP analysis we don't explicitly use the coefficient estimates. Instead, we use these to predict the probability for each category and then use these predictions to make a population estimand. If coefficient estimates, particularly the intercept terms, are overall lower than the true value then this could result in lower predictions than expected.\citep{king2001logistic} The difference between sample mean and model estimate increases as the model grows more complex\cite{puhr2017}. This aligns with rare-events bias (parameters estimates grow smaller relative to truth as the model grows more complex).

\subsubsection*{Impact of measurement constraints}

However, there are other potential causes of this phenomenon. We know that $\pi_i$ is contained in the interval $(0,1)$. Using this, and rearranging equation (\ref{eq:merror})---see Appendix \ref{ssec:proof of constraints} for analytic proof---we find

\begin{equation}
\begin{cases}
        (1 - \gamma)  \leq \widehat{\mbox{Pr}}(y_i^*=1) \leq \delta,& \text{if } \delta > (1-\gamma)\\
        \delta \leq \widehat{\mbox{Pr}}(y_i^*=1) \leq (1 - \gamma) ,& \text{if } \delta < (1-\gamma) \\
        \infty             & \text{if } \delta = (1-\gamma).
 \end{cases}
 \label{eq:merror_constraint}
\end{equation}

\begin{figure}
    \centering
    \includegraphics[width=0.9\linewidth]{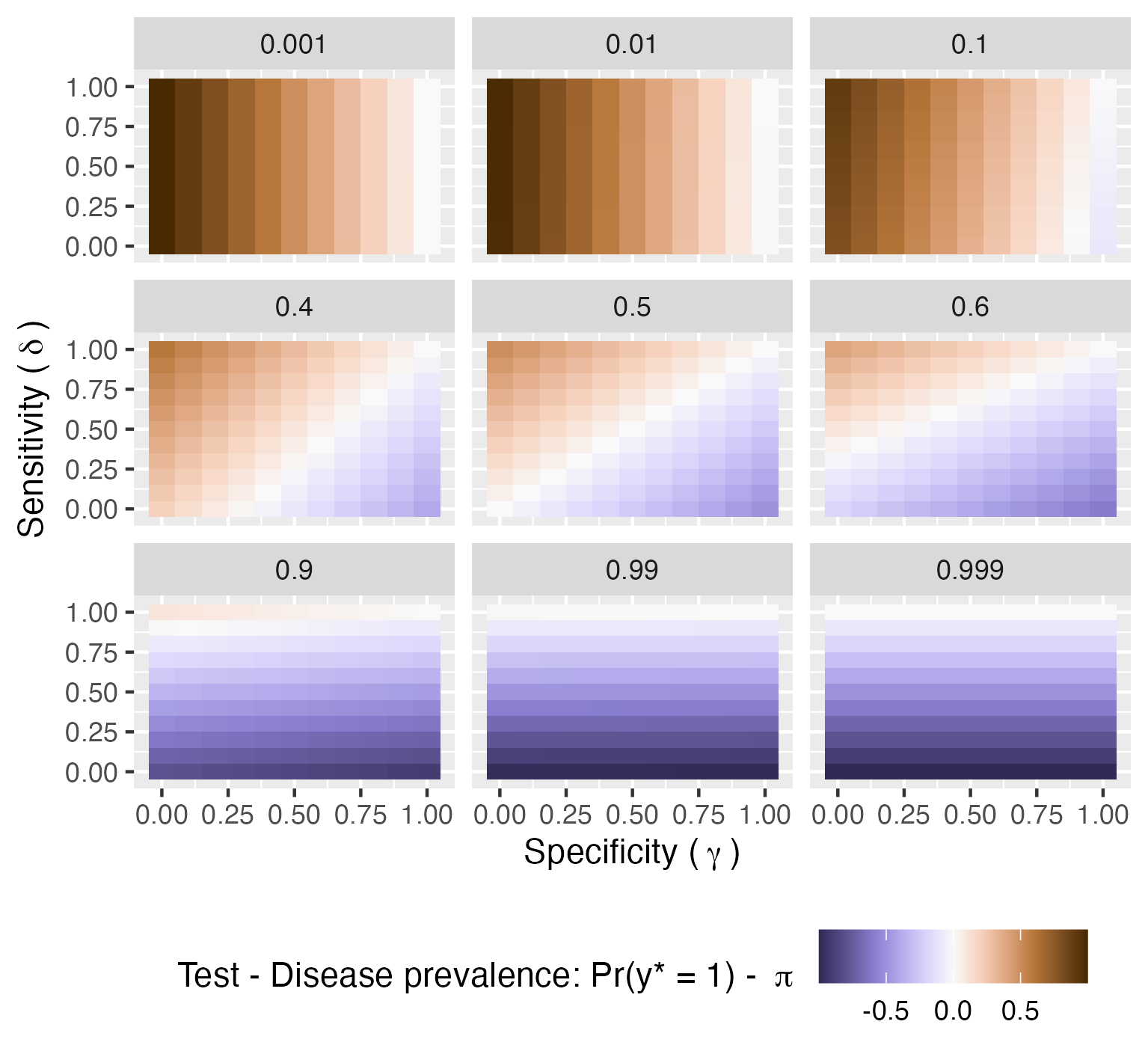}
    \vspace{-.2in}
    \caption{\em Effect of sensitivity and specificity on the accuracy of estimating the true population rate $\pi$ from the observed test prevalence $p$. Orange tones indicate that the observed test prevalence is higher than the disease prevalence, while blue tones indicate observed test prevalence is lower than the disease prevalence. Panels indicate different levels of disease prevalence, demonstrating that when disease prevalence is low, specificity constrains the recovered value.  This shows the lower bound of the inequality. When prevalence is high, the sensitivity bounds the error. In the middle facets, we see a tradeoff between the two, showing the conditional use of inequality.}
    \label{fig:effect_sens_spec}
\end{figure}

\noindent%
The observable test prevalence is constrained to lie between $1 - \text{specificity}$ (the expected rate of positive tests if the population was all negative) and sensitivity (the expected rate of positive tests if the population was all positive) if the specificity is less than $1 - \text{sensitivity}$ and vice-versa if it is greater. 

We demonstrate this visually in Figure~\ref{fig:effect_sens_spec}. We simulate the true prevalence of 9 different levels $(0.001, 0.01, 0.1, 0.4, 0.5, 0.6, 0.9, 0.99, 0.999)$ to show how the true prevalence impacts the relationship between sensitivity, specificity and the observed test prevalence. In the cases where prevalence is very high (bottom row), we see that sensitivity predominately controls the difference between observed and true prevalence (as the difference doesn't change for different levels of specificity). When the true prevalence is very low (top row, most relevant to our example), specificity predominately controls the difference between observed and true prevalence (as the difference doesn't change for different levels of sensitivity). For values in between there is a tradeoff, which provides support for equation (\ref{eq:merror_constraint}). 

These results are important for two reasons. Firstly, because it demonstrates that we could observe similar values of test prevalence for different true disease prevalences--the difference depends entirely on the sensitivity and specificity of the test. Secondly, the relative importance of sensitivity compared to specificity depends on the underlying true prevalence. In our motivating example, we are focused on a rare event, and so we will focus primarily on the impact of estimating specificity (rather than sensitivity). If, however, we are interested in a very common event (or the probability of \textit{not} having COVID-19 antibodies present in the blood), then we would instead focus on sensitivity. We use these results extensively in designing our simulation studies in Experiment II. 

\subsubsection*{Modeling decisions for covariates} 

Bafumi and Gelman \cite{bafumi2007fitting} demonstrate that when fitting a model with an unmodeled coefficient (fixed effects) and a grouping variable modeled as a varying effect, the correlation between the unmodeled coefficient and varying effect can make estimates overly precise. To resolve this, they show  that the group-wise mean of the fixed effect should be included in the model at the group level. However, they only investigate one varying effect and one fixed effect in their simulations. In our work, the models have up to $4$ varying effects (strata, postcode, SEIFA, and age group) and up to $2$ unmodeled coefficients (sex, coded as a binary variable, and SEIFA, coded as a linear predictor). Our modeling decisions also differ from the Bafumi and Gelman paper in that SEIFA is included as both a fixed and varying effect---presumably to capture the anticipated linear effect that increasing SEIFA decreases the prevalence of COVID-19 observed in other studies whilst also allowing for a non-linear effect modeled through the varying effect. It is not clear that these modeling decisions are related to our findings, but for completeness, we do include simulations investigating the impact of unmodeled coefficients.

So far, we have shown evidence that the unexpected results is a model artifact and not a true representation of the data. Due to the complexity of the model, it is not clear what exactly could be causing the model artifact. We identify at least three different potential causes that can interact. In the remainder of this manuscript, we explore each factor independently or in combination with other factors (as necessary) using simulation studies. Following each simulation study, we modify the model fits to real data to see if the simulation study results align with our real data. In doing so, we contribute new knowledge in three different domains before identifying what we believe to be the true cause of the modeling challenge. 

\begin{itemize}
    \item In Experiment I, we demonstrate the implications of prior knowledge when investigating rare events. We showcase the difference in rare-events bias in a Bayesian approach when compared to maximum likelihood.
    \item In Experiment II, we incorporate sensitivity and specificity first as known parameters, then estimate specificity. We demonstrate that when conducting prior predictive checks, measurement error needs to be incorporated. We also highlight the importance of sufficient data in each module of the model. 
    \item In Experiment III, we finish by including varying effects and unmodeled coefficients. We show that even when the varying effects are uncorrelated with the linear predictors, this induces further bias in the estimates. 
\end{itemize}
 
We finish our investigation by demonstrating that, rather than a single cause, the model artifact is caused by the feedback between the measurement and prevalence components of the model. The feedback worsens as more parameters are added to the model, changing the prior predictive expected prevalence. This is then fed through to the specificity estimates, which change as the prevalence model grows more complex. These specificity estimates then change the overall estimate of prevalence during the high reliance on specificity estimates when estimating rare events. At a wider level, this paper contributes

\begin{itemize}
    \item Demonstration of progressive simulation studies to investigate a complex modeling challenge.
    \item Evidence of feedback between measurement model component and prevalence model components, and demonstration that this can be observed in counterintuitive estimates.
    \item Interim recommendations for individuals fitting this particular modeling framework. 
\end{itemize}

\section{Method}
To explore each potential cause of the observed statistical artifact, we use a combination of simulation studies and modified models applied to the motivating example. In this section, we specify the broad simulation design, with modifications on this design described as relevant in each exploration. 

\subsection{Data generation}\label{sec:data_gen}

We adopt a finite population approach, which means that we assume a single, stable population across all simulation iterations for the same experimental condition. The population is created to have $N = 500,\!000$ individuals. Each individual is described by five covariates that represent demographic variables, denoted $X_1,\dots, X_5$. 

Each predictor $X_k$ is first simulated as a continuous variable and then discretized into $L(k)$ groups of equal number of individuals in each group, with a focus on $L(k) = 20$ ($5 \times 20 = 100$ levels) to examine the impact of covariate levels on the estimates. In the motivating example, there are $117$ levels of varying effects that need to be estimated (albeit unevenly across the different parameters $3$ + $102$ + $5$ + $2$ + $5$; we deliberately avoid referring to this as the number of parameters to be estimated, as there is information that is shared across these levels). This leads to a large number of rows for poststratification. Normally, this would be an issue, but as we have shown, the model appears to be the most likely cause of the observed problem. 

We also include the results for a number of covariate levels $L(k) = 4, 10$, and $40$ for Experiments 1.2--2.2 in the appendix to examine the impact of different covariate levels. They generally show slightly more exaggerated bias when the covariate levels increase and the sample size decreases. In Experiment III, we focus only on results for 20 covariate levels for five covariates, as it is the closest to the number of coefficients in our original application (although many more poststratification cells).

The relationship between the probability of the outcome and the covariates was created using the continuous $X$'s instead of the discrete versions to make it easier to control the number of levels in the covariate. 

We define the relationship between the probability of positive antibody response $\pi_i = \text{Pr}(y_i = 1)$ and the demographic variables as
\begin{equation}
     \pi_i = \logit^{-1}(\zeta_0 + \zeta_1X_{1i} + \dots + \zeta_1X_{5i}), \label{eq:pi_y_gen}
\end{equation}
where all covariates have the same strength of relationship to the outcome ($\zeta_1$ is the same value for all covariates). We manipulate the strength of the relationship between the demographic covariates and the outcome. We create two different outcomes in the population. One, $\mathbf{y_0}$, has no relationship to the covariates ($\zeta_1 = 0$); the other, $\mathbf{y_1}$, has a strong relationship ($\zeta_1 = 0.3$). We generate $\mathbf{X_k}$'s from a uniform distribution with lower and upper limits of $-0.5$ and 0.5, respectively, which gives $\mbox{E}(\mathbf{X_k})=0$, so the expected population prevalence is the same for both conditions.

We aim to hold the overall population prevalence steady, given the different relationships between demographic covariates and outcomes. To achieve the prevalence rate of $\pi$, consistent with the original problem, we compute the intercept $\zeta_0$ as,
\begin{equation*}
   \zeta_0 = \logit(  \pi ) - \frac{\zeta_1 (a+b)}{2} K,
\end{equation*}
where $a = -0.5$ and $b = 0.5$ are the lower and upper limits of the uniform distribution, and $K$ is the number of covariates. This reduces to 
\begin{equation}\label{eq:b0}
   \beta_0 = \logit(\pi)
\end{equation}
for both the $\zeta_1 = 0$ and $\zeta_1 = 0.3$ conditions. 

Given the initial $\zeta_0$ value, we then use equation (\ref{eq:pi_y_gen}) to generate the probability of antibody prevalence $\pi_i$ for each individual. We then create the observed probability of positivity test prevalence $\mbox{Pr}(y^*_i=1)$ through equation (\ref{eq:merror}). 

From our prior predictive checks, we see the artifact is present in both sample and population predictions. To simplify the simulation, we use random sampling with the same probability for each unit. We vary the sample size $n$ from 400 to 4000. We repeat each simulation condition 100 times, drawing a new sample for each simulation iteration.

In Experiment 1.1, we examine populations with varying base rates, specifically $\mbox{E}(\pi) = 0.001$, $0.01$, $0.1$, and $0.2$, under the condition that $\zeta_1 = 0$. For all other simulations, we consider only $\mbox{E}(\pi) = 0.01$, which is close to the sample base rate in our motivating example \cite{machalek2022}.

From Experiment 1.2 onward, we have two outcomes for different strengths of the relationship between covariates and outcome. For each experimental setting, we first draw $100$ random samples of two sample sizes of 400 and 4000 from the finite population. Later, in Experiment 2.2 onward, we focus only on sample size equal to 4000 and the strength of the relationship between covariates and outcome at $\zeta_1 = 0.3$. 

\subsection{Estimation method}

Our core manipulation is the impact of adding additional predictors to the model, represented by the sequence of models as described in Table~\ref{tab:models_fit_expdata} (using synthetic data) and Table~\ref{tab:models_fit_realdata} (using the real data).

For each sample, we fit the relevant model. In Experiment I, posterior draws were taken using brms 2.22.0\citep{brms18} with default settings and a backend of \texttt{cmdstanr}\citep{cmdstanr} that uses Stan 2.32.2.\citep{carpenter2017stan} In Experiments II and III, which incorporate sensitivity and specificity, \texttt{brms} could not be used as these models were not included in the set of models coded in that wrapper, so we directly programmed in Stan, adapting the code from our motivating example to ensure consistent prior specification and parameterization. Where used, classical model fits were obtained using glm in the base R stats package.\citep{r2024} Results and visualizations were created using ggplot2 3.5.1.\citep{ggplot16}

\renewcommand{\arraystretch}{2}
\begin{table}[]
\begin{threeparttable}
\resizebox{1.05\textwidth}{!}{
\begin{tabular}{p{0.1cm}p{0.1cm}c>{\centering\arraybackslash}p{1.5cm}>{\centering\arraybackslash}p{2cm}p{2.3cm}p{2.5cm}p{2.6cm}p{2.8cm}}
&& \multicolumn{7}{p{12.6cm}}{\centering \bfseries Investigating three potential issues by simulations} \\
& & & \multicolumn{2}{p{3.5cm}}{\bfseries Experiment I: rare-events bias in logistic models (with covariates)} 
& \multicolumn{2}{p{4.8cm}}{\bfseries Experiment II: Inclusion of the uncertainty of test sensitivity and specificity} 
& \multicolumn{2}{p{5.4cm}}{\bfseries Experiment III: Adding overall effects terms} \\
& & \multirow{4}{*}{\textbf{Model}} & \multicolumn{1}{l}{\textbf{Exp 1.1}} &  \textbf{Exp 1.2} &  \textbf{Exp 2.1} &  \textbf{Exp 2.2} &  \multicolumn{2}{c}{\textbf{Exp 3}} \\
& & & \multirow{2}{=}{\textbf{intercept-only models}} & \textbf{Adding covariates\tnote{1}} & \textbf{Known specificity\tnote{2}} & \textbf{Estimating specificity\tnote{3}} & \multicolumn{2}{c}{\multirow{2}{*}{\textbf{Adding overall effects terms}}}  \\
& & & & & & &  \textbf{One overall effects term}& \textbf{Two overall effects terms}\\
\hline
\multirow{7}{*}{\rotatebox{90}{\textbf{Covariates added sequentially}}} & \multirow{7}{*}{\tikz \draw[<-, thick] (0,4.5)--(0,10);} & \textbf{0} & - & $y \sim 1$ & \multirow{6}{2.5cm}{\vfil The outcome is modified to include measurement error. Sensitivity and specificity are treated as known (not estimated).} & \multirow{6}{2.5cm}{\vfil  Specificity is estimated using a beta prior with varying sample sizes.} & No change & No change \\ 
& & \textbf{1} & - & $y \sim (1 | X_1)$ & & & No change & No change \\
& & \textbf{2} & - & $\ldots + (1|X_2)$  & & & No change & No change \\ 
& & \textbf{3} & - &$  \ldots + (1|X_3)$  & & & \multirow{3}{=}{Overall slope parameter for $X_3$ added} & Overall slope parameter for $X^*_4$ added  \\    
& & \textbf{4} & - &  $\ldots + (1|X_4)$ & & &  & \multirow{2}{=}{Overall slope parameter for $X^*_4$ replaces varying intercept for $X_4$ } \\
& & \textbf{5} & - & $\ldots + (1|X_5)$  & &  & & \\
&  \multicolumn{8}{r}{\tikz \draw[->, thick] (2,1) -- (17,1);\vspace{-5mm}} \\
\multicolumn{9}{c}{\textbf{Built on previous experiments}} 

\end{tabular}
}
\caption{\em In each experiment, covariates are added sequentially from Models 0 through 5. Across the experiments, each variation is added and built on the previous simulations.}
\label{tab:models_fit_expdata}
\begin{tablenotes}
  \item [1] R syntax for fitting the models using \texttt{brm}.
  \item [2] The models are fitted using adapted Stan code compiled using \texttt{cmdstan\_model}.
  \item [3] Sensitivity is treated as known in this experiment.
\end{tablenotes}
\end{threeparttable}
\end{table}

For each model, our estimation procedure is the same. The probability of antibody prevalence for each poststratification cell is estimated by 
\begin{equation*}
    \hat{\pi}_j = \mbox{logit}^{-1}\big(\hat{\beta}_0 + \sum_{k=1}^K\hat{\alpha}^{(k)}_{l[j]}\big).    
\end{equation*}
From this, we estimate the overall antibody prevalence in the population as
\begin{align}
    \widehat{\pi} &= \frac{\sum_{j=1}^J N_j \widehat{\pi}_j}{\sum_{j=1}^J N_j }.
\end{align}
For Experiment I, where sensitivity and specificity are not incorporated into the model, this is the same as estimating $\widehat{\mbox{Pr}}(y_j =1)$. 

\subsection{Evaluation metrics}

The principal metric used in the experiments is bias in our estimate of the overall COVID-19 antibody prevalence :
\begin{equation*}
    \mbox{Bias}_\pi = \hat{\pi} - \pi.
\end{equation*}
\begin{equation*}
\mbox{Bias}_{\beta_0} = \hat{\beta}_0 - \beta_0,
\end{equation*} 
as the rare-events bias literature describes bias in regression coefficients and the intercept term are most likely to affect the overall estimated prevalence. 

In the motivating example, it is not possible to calculate the bias of our estimate as the truth is not known. In the initial prior predictive checks, we compare the difference between estimated disease prevalence and observed test prevalence. As we shall see, this is a relevant factor in interpreting our posterior predictive checks, so we also include metrics comparing the estimate of COVID-19 antibody prevalence ($\hat{\pi}$) against positive test prevalence ($\mbox{Pr}(y^*_i =1)$):
\begin{equation*}
    \Delta_\pi = \hat{\pi} - \mbox{Pr}(y^*_i = 1).
\end{equation*}


\section{Experiments}

\phantomsection 
\section*{Experiment I: rare-events bias}
\addcontentsline{toc}{subsection}{Experiment I: rare-events bias in logistic models (with covariates)}  
\label{sec:rare_events}

Starting with the simplest explanation, we first consider whether the observed increasing bias is caused or related to rare-events bias. This can be explored independently of the estimation of sensitivity and specificity, and for the models in this section, we consider only models of antibody prevalence as if it were measured with no error. 

To explore this factor, we present three elements of evidence. First, we explore whether rare-events bias is observable in a Bayesian modeling framework (a) at all and (b) for the sample size and prevalence rate in our motivating example. Then, we explore the impact of adding additional varying effects into the model, noting previous research that identifies that it is not just the prevalence and sample size that are factors in rare-events bias, but also the number of parameters estimated. Finally, we demonstrate that these simpler models do not replicate the same artifact that was observed in our motivating example. 

\phantomsection 
\subsection*{Experiment 1.1: Intercept-only model}\label{sssec:int-only}
\addcontentsline{toc}{subsubsection}{Experiment 1.1: Intercept-only model}

We first investigate bias in Bayesian logistic models while varying the population prevalence and sample size. We use four conditions to represent the true probability of COVID-19 antibody prevalence within the population. The prevalence is 0.001, 0.01, 0.1, or 0.2. We also vary the sample size $n = 20$, 40, 400, 4000, and 40,000. Rare-events bias has been primarily documented using maximum likelihood estimation \citep{leitgob2013,williams2016}. For this first simulation, we compare maximum likelihood estimation to a Bayesian approach so that if we do not observe rare-events bias in the Bayesian approach, we can confirm that the conditions were sufficient to observe with rare-events bias with maximum likelihood.   

\begin{figure}
\centering
\includegraphics[width=.95\textwidth]{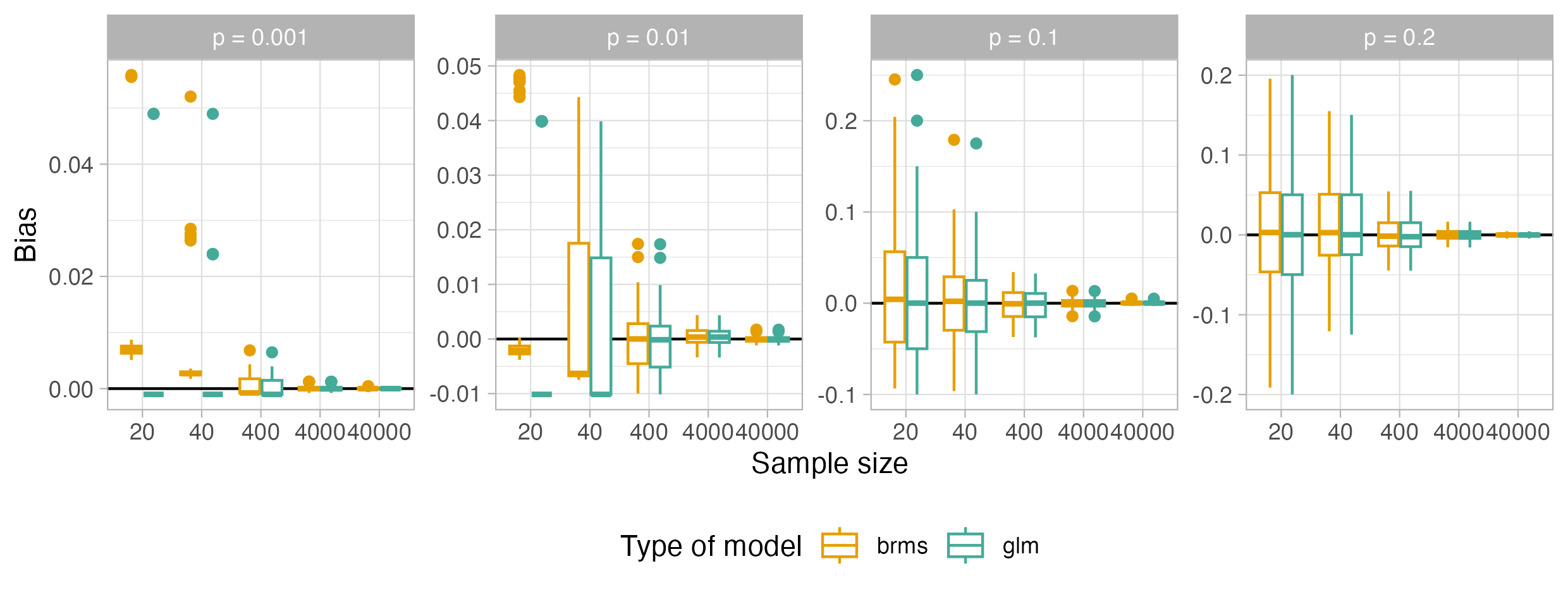}\vspace{-1.5em}
\caption{\em Distribution of bias ($y$-axis) between posterior median estimate for prevalence ($\pi$) and truth over 100 iterations using varying sample sizes ($x$-axis) from the intercept-only models, for true probability of outcome 0.001, 0.01, 0.1, 0.2, in each panel. Box plots are used to demonstrate variance across simulation iterations, with orange used to represent Bayesian models and green used to implement the models implemented using the \texttt{glm} function in R. Negative values indicate where the estimate is smaller than the truth (rare-events bias), while positive estimates indicate where the estimate is larger than the truth. Variance and outliers for smaller sample sizes and low prevalence are driven by the estimates clumping based on sample properties, which is shown more clearly in Appendix \ref{ssec:alliter}.}
\label{fig:int_only_median_est}
\end{figure}

In this first simulation, we are interested only in the presence of rare-events bias. This means that rather than fitting all six models, we fit only an intercept-only model,
\begin{align}
 \nonumber   \mbox{Pr}(y_i = 1) &= \logit^{-1}(\beta_0).
\end{align}
For the Bayesian approach we need to select a prior for the intercept term. To avoid setting the prior based on prior knowledge or simulation study conditions, we instead use the default prior in brms. 
\begin{align}
  \nonumber    \beta_0 &\sim t_3(0, 2.5).
\end{align}
Figure \ref{fig:int_only_median_est} shows interesting differences between the two estimation frameworks. When sample size and prevalence are very small, there is an overall tendency for the generalized linear model to underestimate the true prevalence. Outliers represent iterations that had a positive observed in the sample (all iterations plotted in Appendix \ref{ssec:alliter} to clearly visualize). This demonstrates the previously documented rare-events bias. 

However, when we use a Bayesian estimation method with a relatively uninformative prior, we see that bias is, overall, more positive than the generalized linear model in small sample sizes and rare prevalence. The bias is in the opposite direction, indicating an overestimation of prevalence. As the overall prior for the model puts substantial mass greater than the prevalence in these cases, we can expect that this is due to contributions of the prior when the sample size is small. As expected from the previous literature, neither bias is present when sample sizes and prevalence increase closer to those used in our motivating example ($n=4000$, $p=0.01$). 

The amount of bias shown in rare event estimation in the Bayesian model will depend on the prior chosen, as well as the sample size and true prevalence. In our motivating example, a similarly uninformative prior was chosen. Given that this Bayesian model demonstrates the opposite of rare-events bias, we do not expect that this is the cause of the observed finding. However, to complete our investigation of this, we next explore the impact of a greater number of covariates in the model. As there is a deviation in bias between the maximum likelihood and Bayesian implementations, we use only the Bayesian implementation going forward.

\subsection*{Experiment 1.2: Adding covariates}
\addcontentsline{toc}{subsubsection}{Experiment 1.2: Adding covariates}

We next investigate the impact of additional covariates on the model in terms of rare-events bias. To explore this, we sequentially add covariates to the model as described in Table~\ref{tab:models_fit_expdata}. 

\begin{figure}
\centering
\includegraphics[width=.8\textwidth]{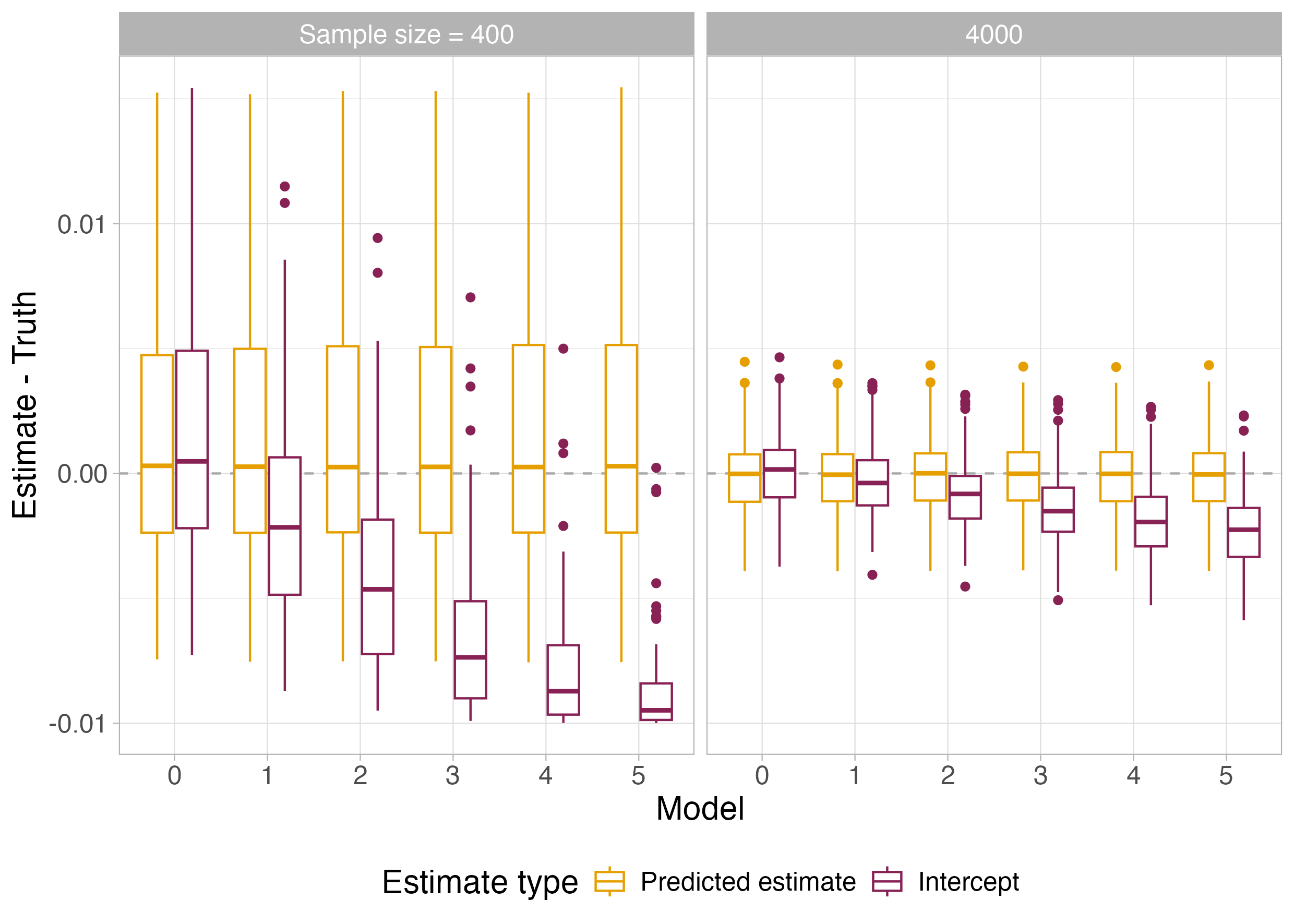}
\vspace{-1em}
\caption{\em Bias for estimates ($y$-axis) using models 0--5 with sequentially added covariates ($x$-axis) for 100 iterations when $\zeta_1 = 0.3$, sample size $= 400$ (left panel) and 4000 (right panel). Box plots are used to demonstrate variance across simulation iterations, with orange used to represent the predicted sample estimates, and purple used to represent the estimated intercept terms. The predicted estimates are largely unbiased across models, while the estimated intercept shows increasing negative bias as the model has more covariates. The pattern is less exaggerated when the sample size is $4000$.}
\label{fig:with_cov_n20}
\end{figure}

The left panel of Figure \ref{fig:with_cov_n20} shows that when the sample size equals to 400, adding covariates sequentially does not change the estimated prevalence of COVID-19 antibodies (orange box plots) across models. However, bias appears in the intercept (purple) as the number of covariates increases, a pattern that becomes less pronounced when the sample size is 4000 in the right panel. See Appendix \ref{ssec:covbeta0.3} for results on other numbers of covariate levels. This trend persists whether there is a strong relationship between the covariates and the outcome ($\zeta_1 = 0.3$) or when there is no relationship ($\zeta_1 = 0$), as shown in the appendix. Investigation of the other parameter estimates ($\beta_1$ through $\beta_5$) suggests that they are biased in the opposite direction to the intercept, which leads to the overall lack of change in the population prevalence estimate. 

From this, we can conclude that, while we do observe bias in the estimated intercept parameter, the overall poststratification estimate does not change as the model grows more complex. 

\subsection*{Real data}
\addcontentsline{toc}{subsubsection}{Real data}

\begin{figure}
\centering
\includegraphics[width=0.9\textwidth]{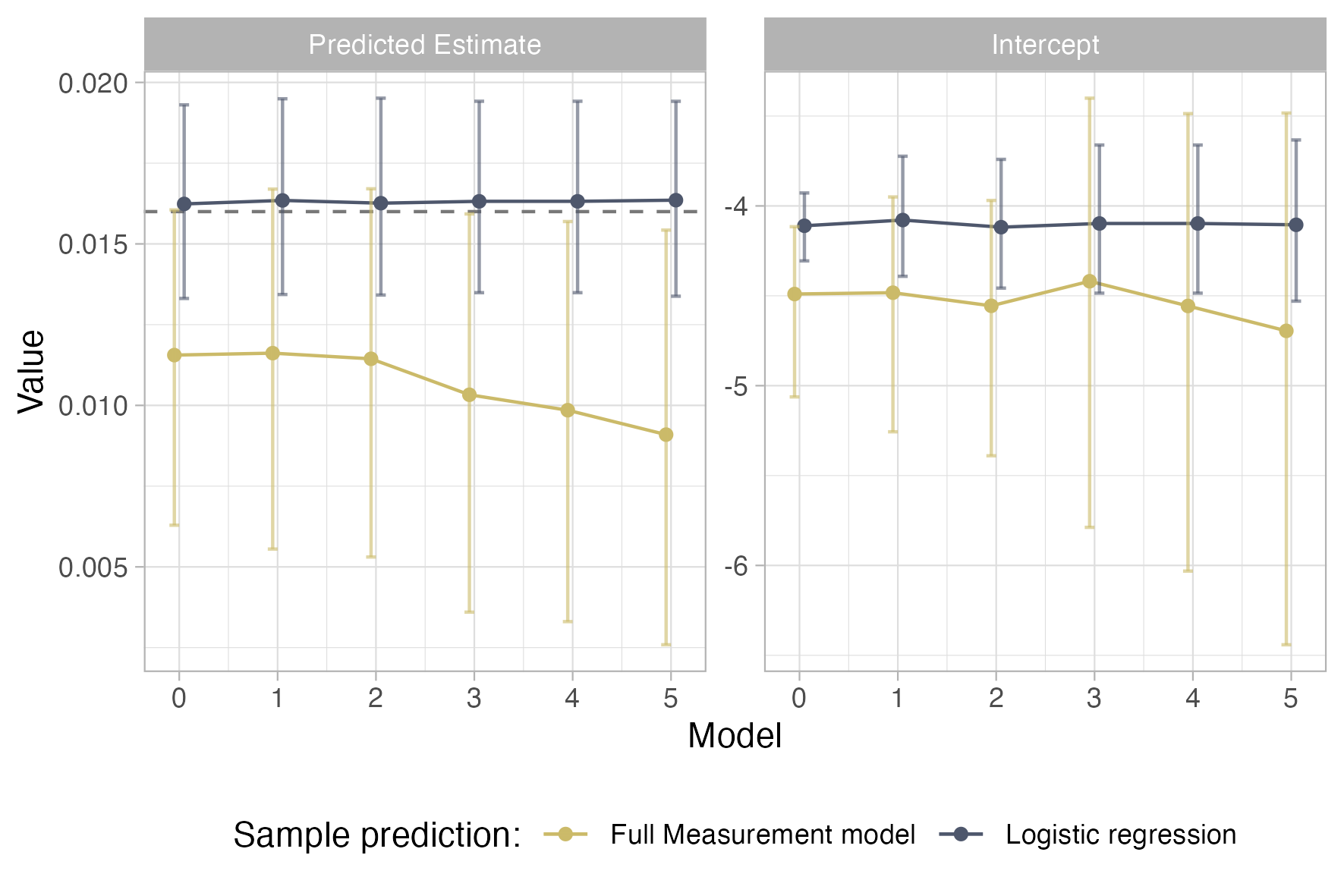}
\vspace{-1.3em}
\caption{\em Estimate for sample mean and intercept parameter ($\beta_0$) from models 0--5 fit with the real data. The black line shows the predicted estimate without measurement error and overall effects terms, and the yellow line shows results from the full model with measurement error and overall effects terms. The right panel plots the intercept for these two models, while the left panel shows the prevalence estimate for the sample. The models that do not include measurement error do not exhibit underestimation or a downward slope as additional covariates are added. This suggests that the issue may stem from the inclusion of either measurement error or unmodeled coefficients in the model.}
\label{fig:og_int}
\end{figure}

To confirm that rare-events bias would not explain the artifact documented in our motivating example, we apply the same model (without measurement error) to the application data. Figure~\ref{fig:og_int} plots the original model estimates and compares them to those obtained from a model without the estimation of sensitivity and specificity included in the model. As expected from the simulation study, there is little evidence of bias in either the overall estimate or the intercept term. Rare-events bias could have contributed to increasing bias as the model grew more complex, particularly in the intercept term, but the sample size and underlying prevalence are sufficient to avoid this.

\section*{Experiment II: Inclusion of uncertainty in test sensitivity and specificity} \label{sec:sens_spec}
\addcontentsline{toc}{subsection}{Experiment II: Inclusion of the uncertainty of test sensitivity and specificity}

We have presented evidence to suggest that model behavior is unlikely to be caused by rare-events bias. We now move to exploring the more complicated model. There are two contributory components we wish to investigate. The first is the impact of the inclusion of sensitivity/specificity at all: as specificity is the limiting factor for low prevalence cases, sensitivity is treated as known in the results described in the following sections.  The second component of the model is the estimation of sensitivity/specificity to propagate uncertainty through to the final estimate of prevalence. In Experiment 2.1, we focus just on the inclusion of sensitivity and specificity into the model, and then in Experiment 2.2, we add in an estimation component. 

\subsection*{Experiment 2.1: Known specificity}
\label{sec:fix_sens_spec}

Our first investigation aims to determine whether accounting for sensitivity and specificity impacts the recovery of the underlying COVID-19 antibody prevalence. In Figure~\ref{fig:effect_sens_spec}, we see that for lower base rates, specificity has a greater impact on the comparison of the observed sample rate to the true population rate. With this in mind, we focus specifically on varying specificity at 0.98, 0.99, 0.995, and 1.00,  holding the sensitivity steady at 1.00. When interpreting the results, it is helpful to remember that our motivating example used a measurement with an estimated specificity of $797/800 = 0.996$. As with previous analyses, we sequentially add covariates into the model. We record the predicted estimates and intercept across all 100 iterations.

\begin{figure}
\centering
\includegraphics[width=0.9\textwidth]{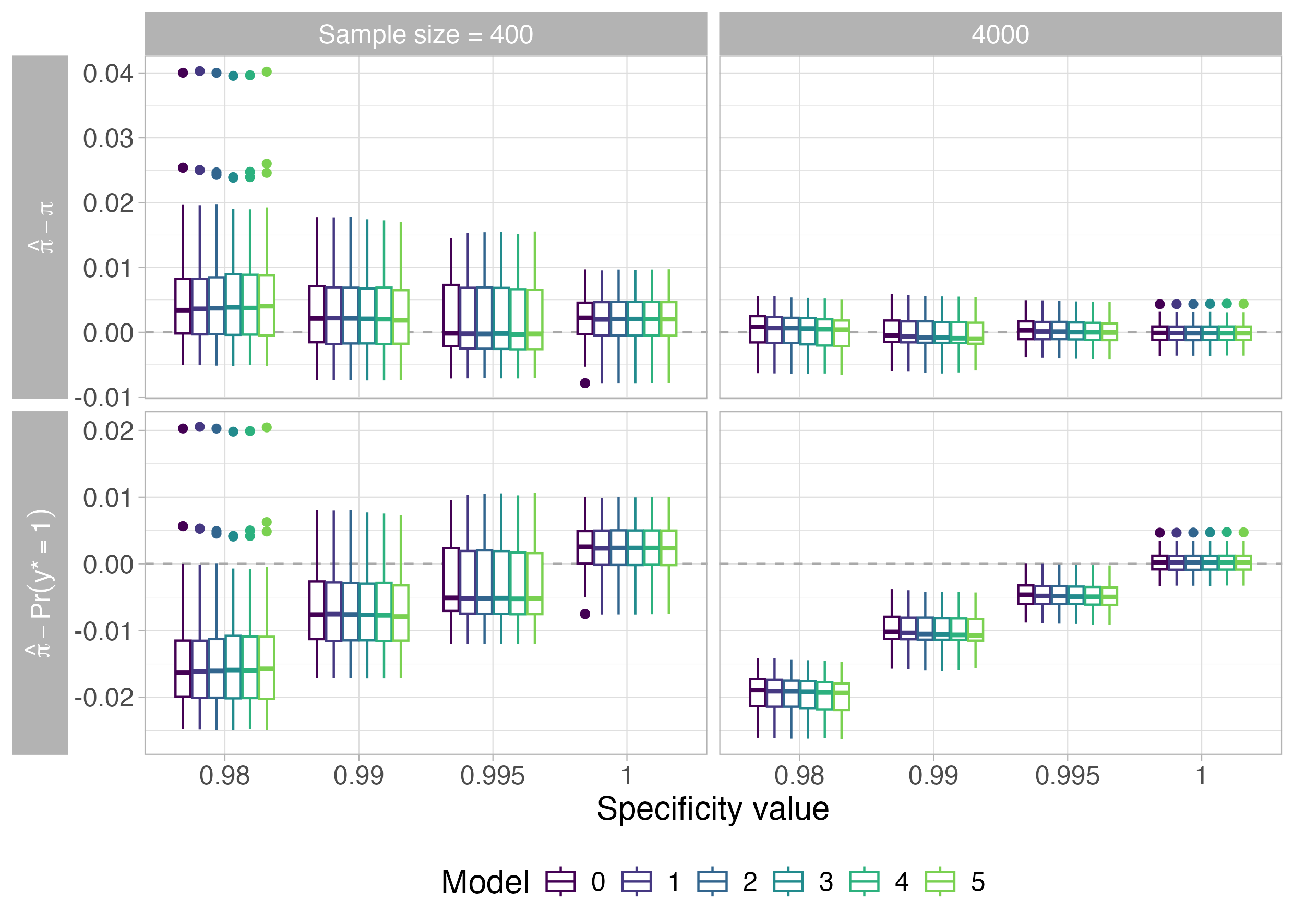} 
\setlength{\abovecaptionskip}{-6pt}
\caption{\em Difference between estimate of COVID-19 antibody prevalence and the true COVID-19 antibody prevalence (top panels) in comparison to the difference between estimate of COVID-19 antibody prevalence and true positive test prevalence (bottom panels). Covariates are sequentially added for each model (color, side-by-side box plots) for different specificity values ($x$-axis). A sample size of 400 (left panels) and 4000 (right panels) with 20 covariate levels (results for other covariate levels are included in Appendix~\ref{ssec:noEstSpec_app}). While the estimate is relatively accurate for true COVID-19 prevalence (particularly with a larger sample size), when compared with test prevalence, the estimate is impacted by specificity. }
\label{fig:noEstSpec}
\end{figure}

Figure \ref{fig:noEstSpec} shows the difference between the estimated COVID-19 antibody prevalence in the population and the true COVID-19 antibody prevalence in the population (top panels). The difference between the estimated COVID-19 antibody prevalence in the population and the true prevalence of positive tests in the population is also shown (bottom panels). The shows see a vertical shift upwards as specificity varies when estimates of COVID-19 antibody prevalence are compared to test prevalence, but relatively unbiased estimates when compared to the true COVID-19 antibody prevalence in the top panel. 

\begin{figure}
\centering
\includegraphics[width=.7\textwidth]{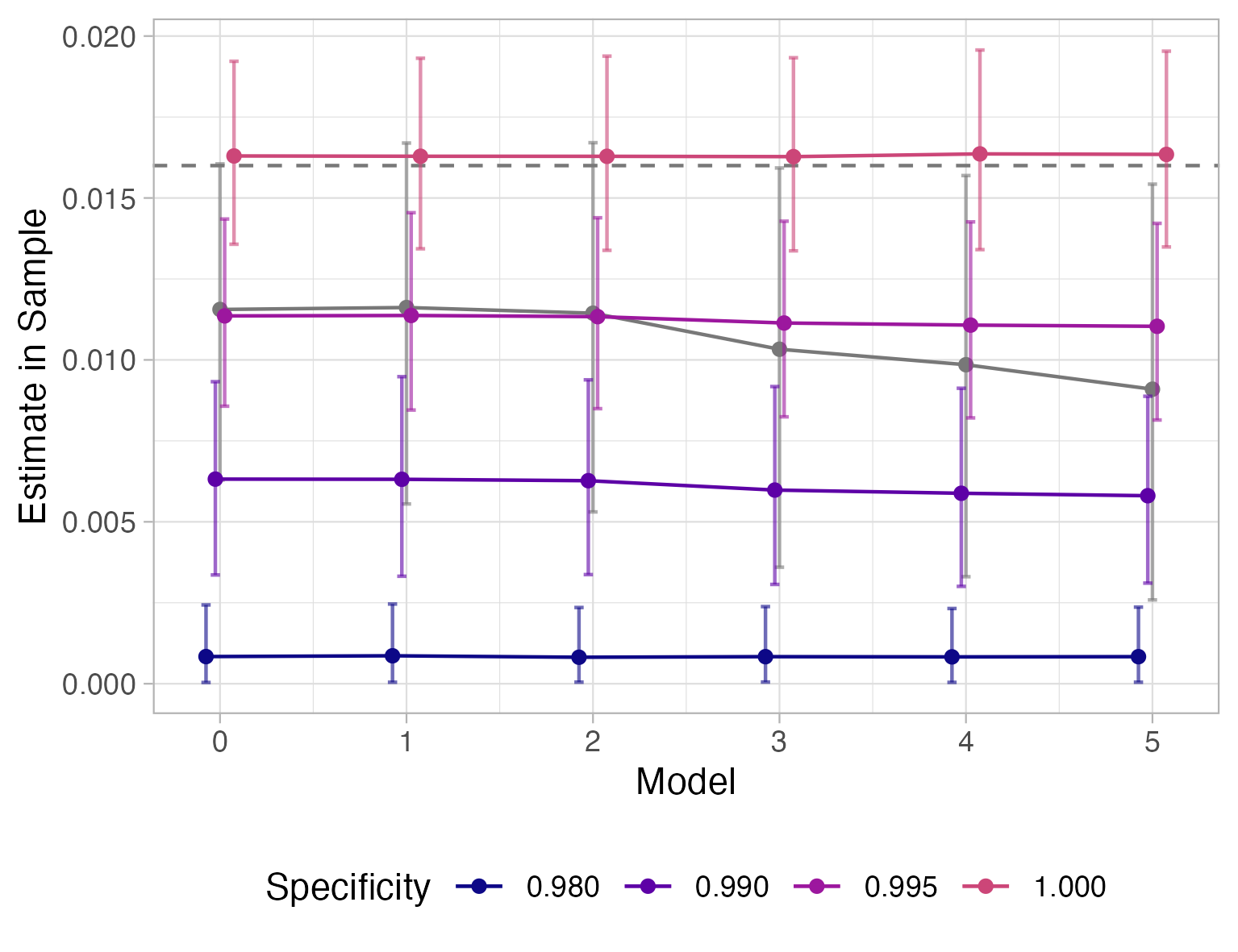}
\setlength{\abovecaptionskip}{-6pt}
\caption{\em Estimate of sample mean with real data using models 0--5 with different levels of measurement error. As we vary the specificity value (dark shade = smaller specificity, lighter shade = higher specificity), we see the overall estimate for all models shift upwards, but no change as the models grow more complex. The gray solid line shows the original model estimates, which align for the first three models with the specificity value of $0.995$, the closest value to the specificity estimated in the real data example. This suggests the shift downward from the sample mean (gray dotted line) is attributable to accounting for measurement error; however, the increasing slope as the model grows more complex is not.}
\label{fig:noEstSpecRealData}
\end{figure}

We then apply this model to the real-world data. This model has two major differences from the originally applied model in the motivating example. Overall effects are not included in the model at all, and sensitivity/specificity are not estimated. We use this model and compare it to the results of Figure~\ref{fig:noEstSpec}.  Figure~\ref{fig:noEstSpecRealData} shows the error of comparing a posterior predictive check of the estimated COVID-19 prevalence to the sample test prevalence. In the motivating example, we see the estimate of population prevalence decrease as the specificity decreases. When compared to the sample mean, this suggests a negative bias. However, this is driven by the differences in comparison to the sample mean (the observed COVID-19 antibody test prevalence) and the true population base rate (the true occurrence of COVID-19).  

This suggests that the overall shift downward from the sample mean is driven by the specificity component in the model, and most likely still a good adjustment from the overall sample mean. When the specificity is closest to the specificity used in our motivating example ($0.995$, light purple), the initial model estimates (with no overall effects and specificity/sensitivity estimated) align very closely. When the specificity is $1.0$ (light pink), the overall predictive estimate is very close to the sample mean. 

\subsection*{Experiment 2.2: Specificity estimated from data} 
\label{sec:est_sens_spec}
\addcontentsline{toc}{subsubsection}{Experiment 2.2: Specificity estimated from data}

Varying the specificity moves the overall estimate of the model, but previous results suggest that this is driven by differences in the inferred overall population COVID-19 antibody prevalence when compared to the observed test prevalence. In our simulation study, we observed relatively small amounts of bias in estimating the true prevalence with a sample size of $4000$, suggesting that this shift is moving toward a correct estimate of population prevalence. 

In this experiment, we focus specifically on adding in the estimation of specificity within the model. As in Experiment 2.1, we consider four specificity values but vary the associated sample size used to estimate these values, setting them at $m_\gamma = 400$, $800$, $1200$, and $8000$. These values can be compared to 800, the amount of test assay data in the motivating example of Machalek et al.\citep{machalek2022} The other levels of data were chosen to vary the information whilst making it possible to retain exactly the same sample estimate of specificity. Table~\ref{tab:spec} shows the four different specificity values and four different sample sizes, and the corresponding false positive and true negative rates. The values closest to the one used by Machalek et al.\ (3 false positives and 797 true negatives) are highlighted. We again model only the specificity and set the sensitivity as 1. All other modeling decisions are identical to Experiment 2.1. 

\renewcommand{\arraystretch}{1.1}
\begin{table}
\centering
\begin{tabular}{r rrr rrr rrr rrr}
 && \multicolumn{2}{c}{Spec = 0.98} &&  \multicolumn{2}{c}{Spec = 0.99} && \multicolumn{2}{c}{Spec = 0.995} && \multicolumn{2}{c}{Spec = 1} \\
 $m_{\gamma}$ && FP & TN && FP & TN && FP & TN && FP & TN \\ \hline
400  && 8 & 392  && 4 & 396 && 2 & 398 && 0 & 400  \\
800  && 16 & 784 && 8 & 792 && \cellcolor{lightgray}4 & \cellcolor{lightgray}796 && 0 & 800  \\
1200 && 24 & 1176 && 12 & 1188 && 6 & 1194 && 0 & 1200 \\
8000 && 160 & 7840 && 80 & 7920 && 40 & 7960 && 0 & 7840 
\end{tabular}
\caption{\em Sample size and corresponding expected false positive (FP) and true negative (TN) counts for calibration data corresponding to different calibration sample sizes $m_{\gamma}$ and different values of specificity, {Spec  = E(TN)/(E(TN) + E(FP))}.}
\label{tab:spec}
\end{table}

\begin{figure}[!htb]
\centering
\includegraphics[width=0.9\textwidth]{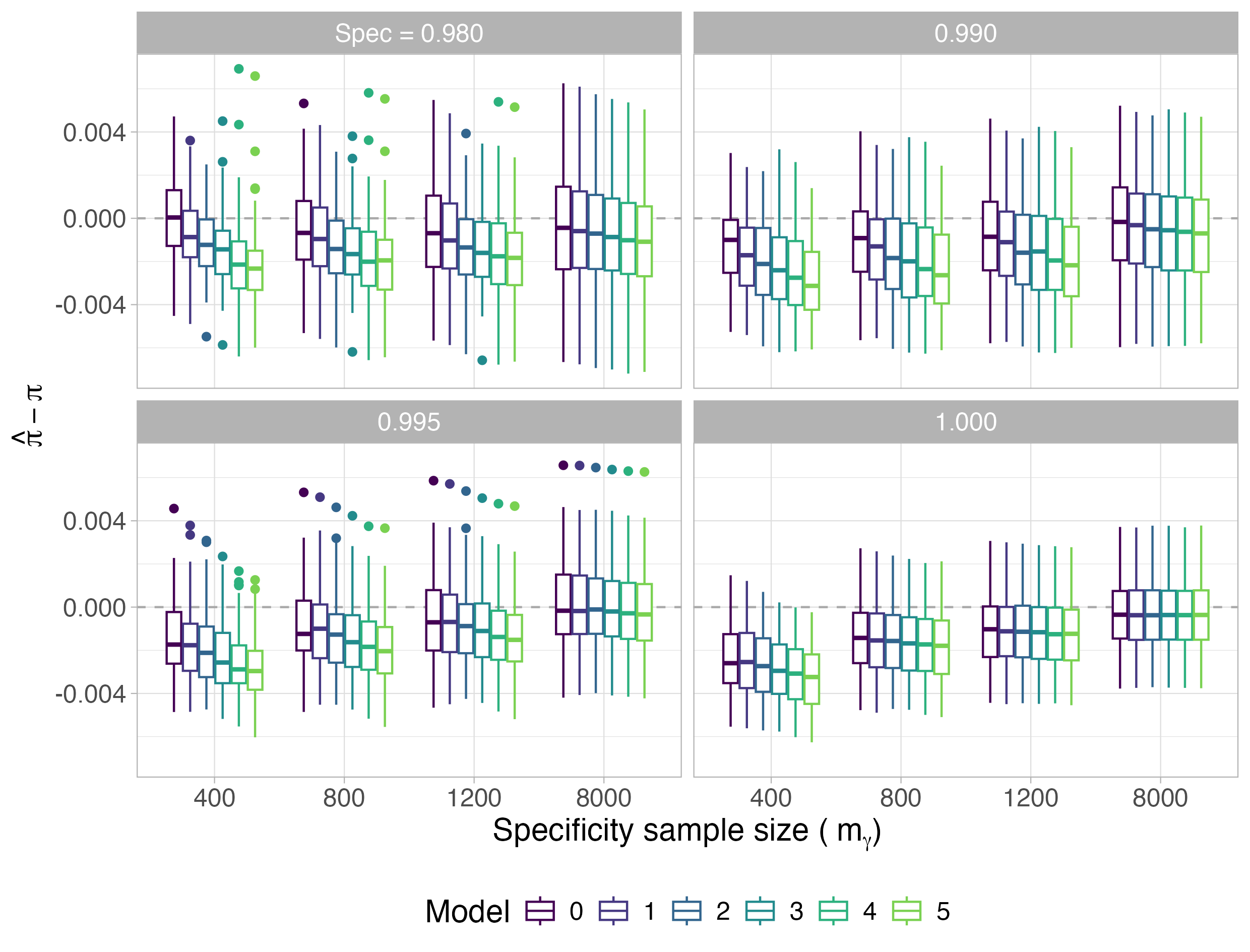}
 \setlength{\abovecaptionskip}{-5pt}
\caption{\em Bias ($y$-axis) between the estimated COVID-19 antibody prevalence and true COVID-19 antibody prevalence using models with sequentially added covariates (color). The sample size $m_\gamma$ of the assay calibration data used to estimate specificity is represented on the $x$-axis. Models presented were fit with a sample size of 4000 and 20 covariate levels for each covariate. The panels represent different values of specificity. The downward trend as the models grow more complex is larger when smaller sample sizes are used to estimate specificity. The comparable condition to the motivating example is sample size = 800 and specificity  = 0.995.}
\label{fig:estSpec}
\end{figure}

When incorporating the estimation of specificity into the model, Figure~\ref{fig:estSpec} shows a negative bias, rather than a slight positive bias as previously seen in Figure~\ref{fig:noEstSpec}. The shift downward is most striking in smaller sample sizes and, surprisingly, when true specificity is closer to 1. The relationship with sample size suggests that this is potentially an impact of the model prior. However, the $\mbox{beta}(TN,FP)$ distribution is centered on the specificity rate regardless of sample size, suggesting potential feedback through the model. Our simulation results show a decreasing trend as the model grows more complex, further strengthening this argument (as increasing the number of parameters in the model also changes the prior). This suggests that our ability to infer the true COVID-19 prevalence might worsen (slightly) as the model becomes more complex. The bias appears worse overall when specificity is higher. 

\begin{figure}[!htb]
\centering
\includegraphics[width=0.9\textwidth]{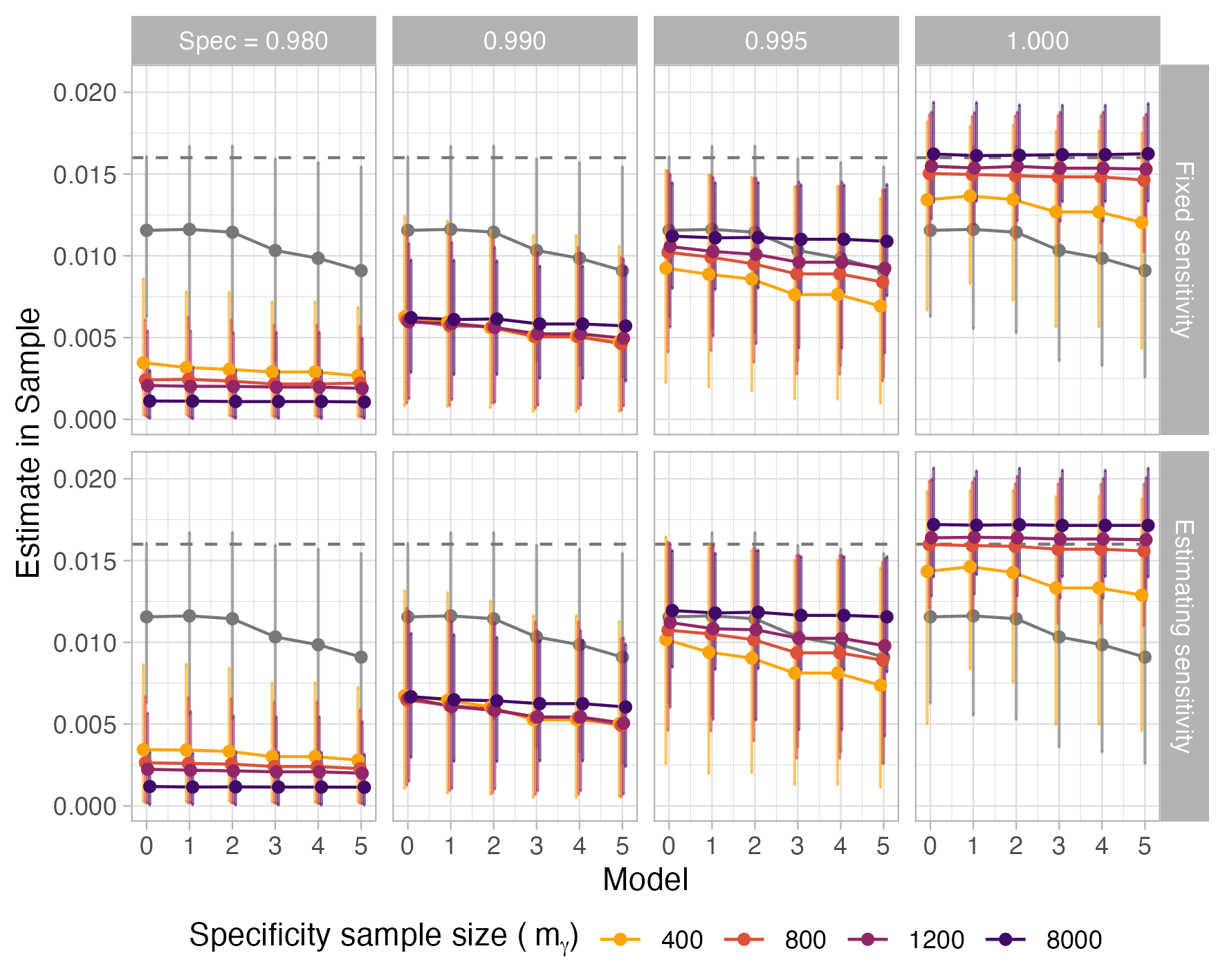}
    \setlength{\abovecaptionskip}{-5pt}
    \caption{\em  COVID-19 antibody estimates ($y$-axis) for models 0--5 ($x$-axis) with specificity estimated. Each panel represents a different specificity $(0.980, 0.990, 0.995, 1.000)$. Colors indicate different sample sizes used to estimate specificity, with darker hues representing larger sample sizes and lighter hues representing smaller sample sizes. The original model estimates are marked in gray, with the dotted line indicating the sample proportion of positive tests. The sample used to estimate COVID-19 antibody prevalence was held constant at $4000$.}
    \label{fig:estSpec_real}
\end{figure}

In Figure~\ref{fig:estSpec_real}, we estimate the COVID-19 prevalence using models that incorporate specificity uncertainty. We see a number of interesting trends. Like before, we see an overall shift based on the underlying specificity (all lines move vertically across the facets). 

However, when we incorporate uncertainty, we see a greater effect of the sample size used to estimate specificity, and an interaction in how this is observed with the value of specificity. When specificity is near 1 and estimated with a large sample size, the estimate of prevalence is roughly the sample prevalence of positive tests, which aligns with previous findings; see Figure~\ref{fig:noEstSpecRealData}. Decreasing the sample size used to estimate specificity results in a lower prevalence. This effect is reversed when the specificity is lower, with a smaller sample size resulting in a higher underlying prevalence.  

We also see an impact of the number of parameters included in the model. As the model becomes more complex, the estimates decrease. This is particularly interesting, as it suggests that less data to inform the estimation of specificity results in a greater susceptibility to changing the prior of the multilevel component of the model (which happens as more parameters are added). 

\vspace{-1.2em}

\section*{Experiment III: Adding non-varying effects terms}
\addcontentsline{toc}{subsection}{Experiment III: Adding overall effects terms}

In Experiments I and II, we have focused on models with just varying effects. This allowed us to explore the performance of a standard MRP method with rare-event (Experiment I) and when incorporating sensitivity and specificity estimation (Experiment II). However, in the original methodology used by Machalek et al., non-varying (fixed/overall) and varying effects were both used. 

In the original models, SEIFA was included as both an non-varying and varying effect. This can be useful when there is a strong correlation between the predictors and group effects.\citep{bafumi2007fitting} SEIFA scores are socioeconomic quintiles measured at the postcode-level. Since we are also modeling postcode as a varying effect, this high correlation between SEIFA and postcode may compromise the estimates. This means that from model 3 onward, we explored the encoding of both an non-varying and varying effect for the same parameter. 

Sex, a two-level variable, was also included in the variable as a non-varying effect term. This is because, with just two levels, it is impossible to estimate the variance term on a varying effect.\citep{gelman2006} This variable is included in model 4 onward. In the following experiment, we code and model our simulated variables in a similar manner to that of our motivating example to explore whether this is the third contributing factor. 

To do this, we use the continuous version of $X_3$, denoted as $X_3^*$, which we will use in the model alongside its corresponding discretized variable, which will still be incorporated as a varying effect as SEIFA is in the motivating example.\citep{bafumi2007fitting, gelman2020} Next, we add a second binary variable $X_4^*$ that is derived by discretizing the continuous $X_4$ in the population into $L(k) = 2$ groups. 

Equation~(\ref{eq:oneFE}) provides an example of how model 3 has been modified for this simulation. In addition to an overall intercept and three varying effects, an additional overall slope parameter is added for $X^*_3$, the continuous version of $X_3$. To be consistent in investigating the effect of two non-varying effects terms from model 3 onward, we also add an overall slope parameter for $X^*_4$ to model 3 to investigate the effect of two non-varying effects (equation~(\ref{eq:twoFE})). In the original application, model 3 only has one non-varying effects term.

For models 4 and 5, the slope parameter for $X^*_4$ replaces the varying intercept for $X_4$ from earlier experiments. This is to remain consistent with the original application, where the binary variable is modeled as a non-varying effect instead of a varying intercept term. This means that in model 4 it would be the same as in equation (\ref{eq:twoFE}). See Table \ref{tab:models_fit_expdata} again for the adaptation to different experiments. 
\begin{align}
\pi_i &= \logit^{-1}(\beta_0 + \beta_1 X_3^* + \alpha_{X_{1[i]}}^{X_1} + \alpha_{X_{2[i]}}^{X_2} + \alpha_{X_{3[i]}}^{X_3}) \label{eq:oneFE} \\ 
\pi_i &= \logit^{-1}(\beta_0 + \beta_1 X_3^* + \beta_2 X_4^* + \alpha_{X_{1[i]}}^{X_1} + \alpha_{X_{2[i]}}^{X_2} + \alpha_{X_{3[i]}}^{X_3}). \label{eq:twoFE}
\end{align} 

\begin{figure}
\centering
\includegraphics[width=0.9\textwidth]{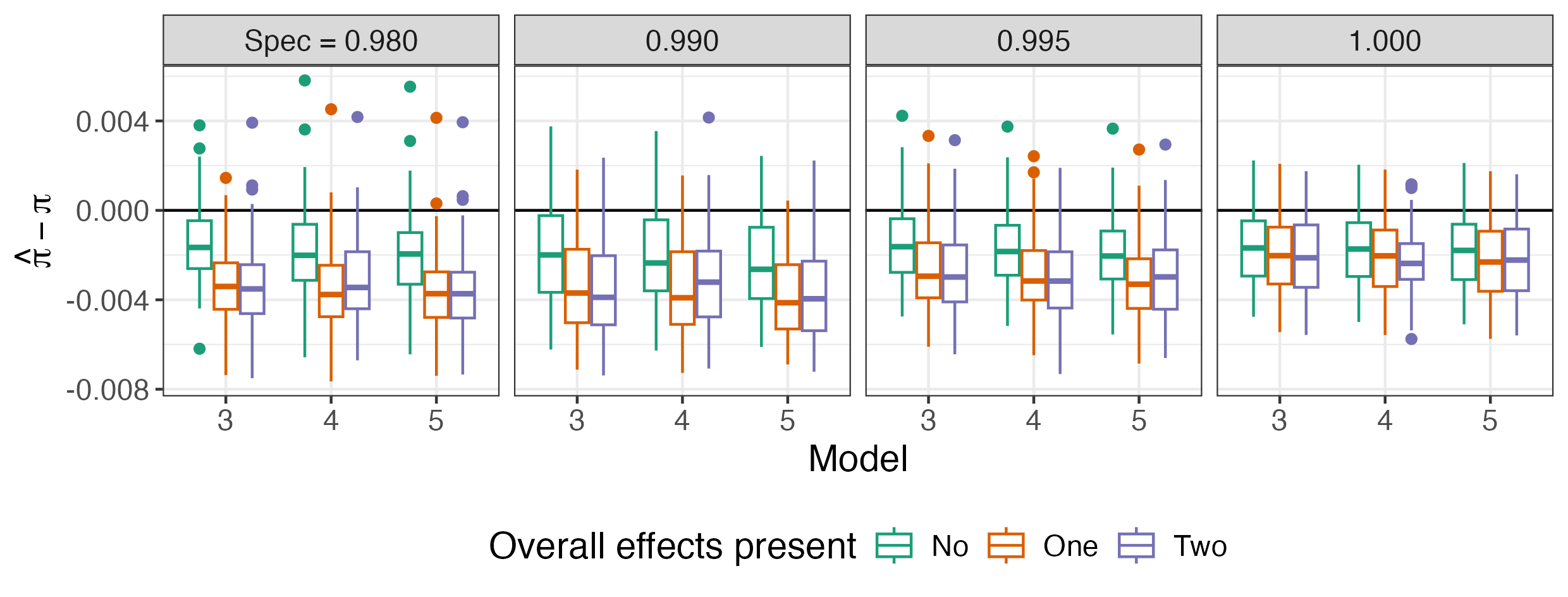} 
 \setlength{\abovecaptionskip}{-5pt}
\caption{\em Difference in estimated and true prevalence of models  ($y$-axis) with different numbers of overall  effects. A sample of 4000 for the primary data and a sample size of 800 is used to estimate specificity. Covariates have 20 levels. Box plots represent variation across 100 iterations. We add either one or two overall slope coefficients (fixed effects) into Model 3 onward, which we focus on for this plot. The horizontal panels represent changing specificity value.}
\label{fig:diff_m0}
\end{figure}

We run the models accordingly and record the predicted estimate. To better illustrate the differences between models, we calculate the bias of the predicted estimates by comparing them to those from model 0. Figure \ref{fig:diff_m0}  shows the difference in the bias of predicted estimates relative to model 0. See Appendix \ref{ssec:exp3_app} for other results, as the increasing prior sample size $m_\gamma$ and the specificity value approaches 1, the bias gradually diminishes.

\begin{figure}
\centering
\includegraphics[width=0.8\textwidth]{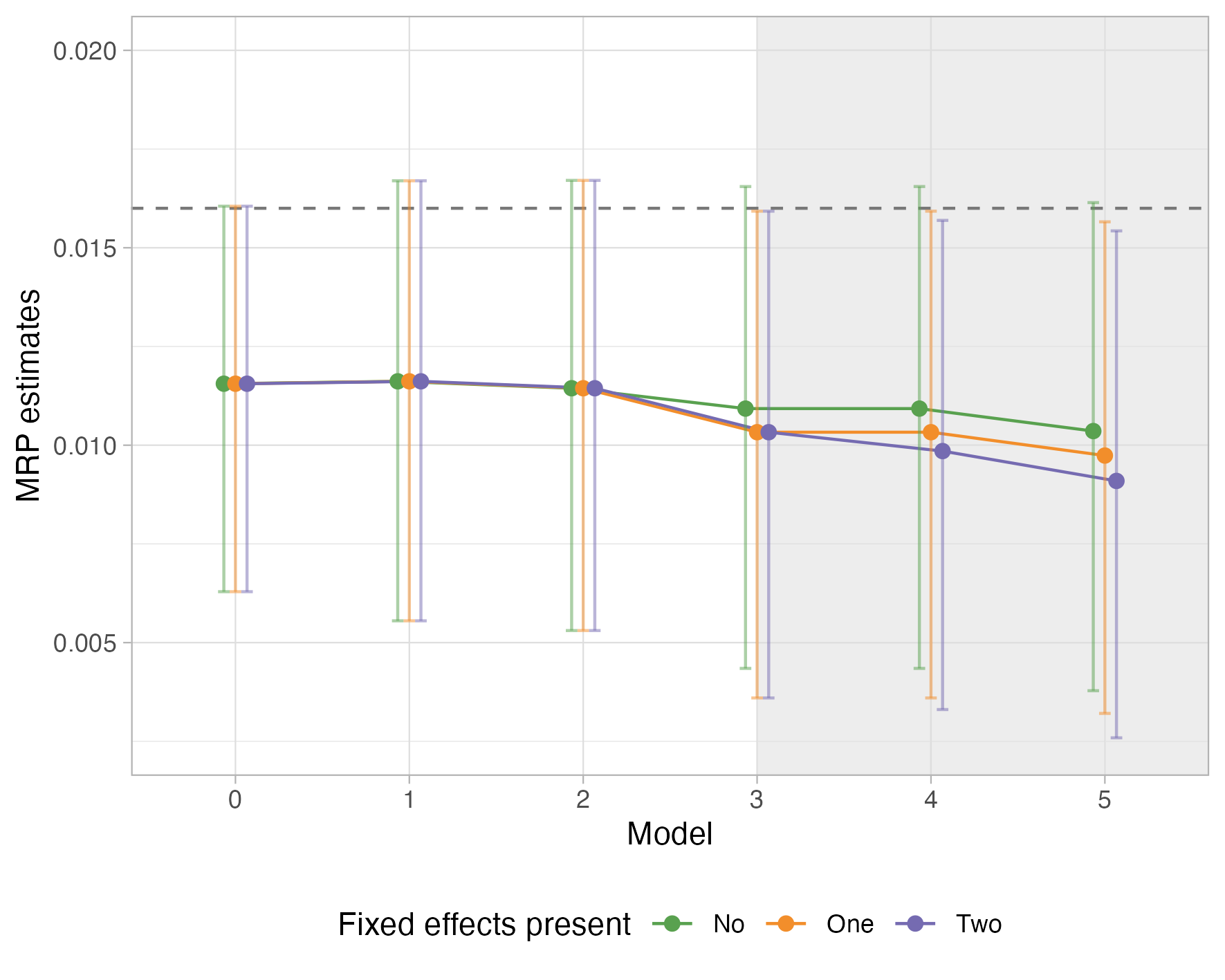}
\vspace{-.2in}
\caption{\em Predicted estimates from models 0--5 with measurement error estimation, holding sensitivity fixed at 1 and a different number of non-varying (fixed) effect terms on the original data. Predictors are only added from model 3 onward: green indicates removal of both, orange is where we keep one, and purple represents the original model estimates with one predictor for model 3 and two predictors for models 4 and 5. The plot shows that without additional predictors, the model estimates show less of a downward trend.}
\label{fig:og_withoutFE}
\end{figure}

In the original data application, non-varying effect terms are introduced starting from Model 3. We remove them one at a time to assess their impact. Figure \ref{fig:og_withoutFE} demonstrates that, in their absence, the estimates remain more stable across models, with a smaller decrease compared to the previous model.

\section{Feedback between components of the model} 

In the three preceding experiments, we have demonstrated that it is not one single issue that is the culprit for this phenomenon, but rather an interaction between the observed prevalence of positive tests, the inclusion of specificity and the estimation of specificity. In particular, we note the impact of the amount of data to estimate the specificity parameter ($m_\gamma$) on the estimates for the population COVID-19 antibody prevalence. 

This indicates feedback in the model, where the mixed effects component impacts the specificity, particularly when there is less data to estimate specificity. In this final analysis, we focus specifically on the interaction between the overall model prior, model complexity and parameter estimation. 

\begin{figure}
\centering
\includegraphics[width=0.8\textwidth]{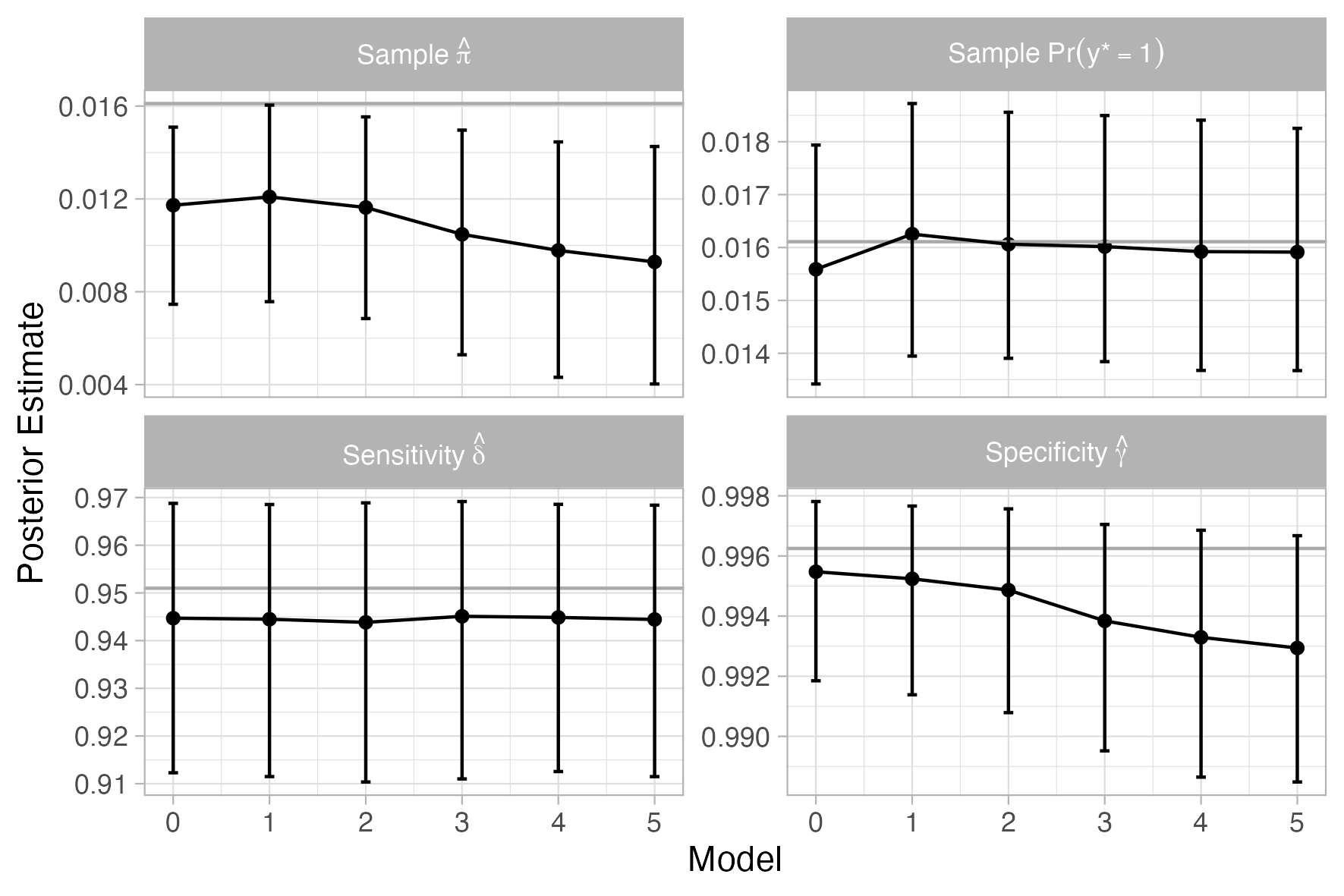}
\caption{\em Posterior estimates for the population prevalence $\hat{\pi}$, sample prevalence $\widehat{\mbox{Pr}}(y^*=1)$, sensitivity $\hat{\delta}$, and specificity $\hat{\gamma}$. Points represent posterior median, while uncertainty intervals indicate 80\% credible intervals. Grey lines indicate observed sample prevalence in the top panel or observed sample sensitivity and specificity, respectively, in the bottom panels. }
\label{fig:real_example_sens_spec_parameters}
\end{figure}

Focusing on real data in Figure~\ref{fig:real_example_sens_spec_parameters}, we see the posterior predictive distributions for population prevalence and test prevalence in the top panels. The population prevalence is both shifted downward and decreases as the model grows more complex. We can also see that this is not observed in our estimates of $\mbox{Pr}(y^* = 1)$, suggesting that the decrease in the model is due to other components in the model. 

In the bottom panels, we plot the posterior estimates for sensitivity and specificity. While sensitivity remains constant for all models, specificity shows a decrease, mirroring the decrease in the population prevalence. This is interesting as the information to estimate specificity (true negative and false positive rate) does not change across these models. What does change across these models is the overall prior for the model---or the prior predictive distribution for the population estimate. 

\begin{figure}
\centering
\includegraphics[width=0.9\textwidth]{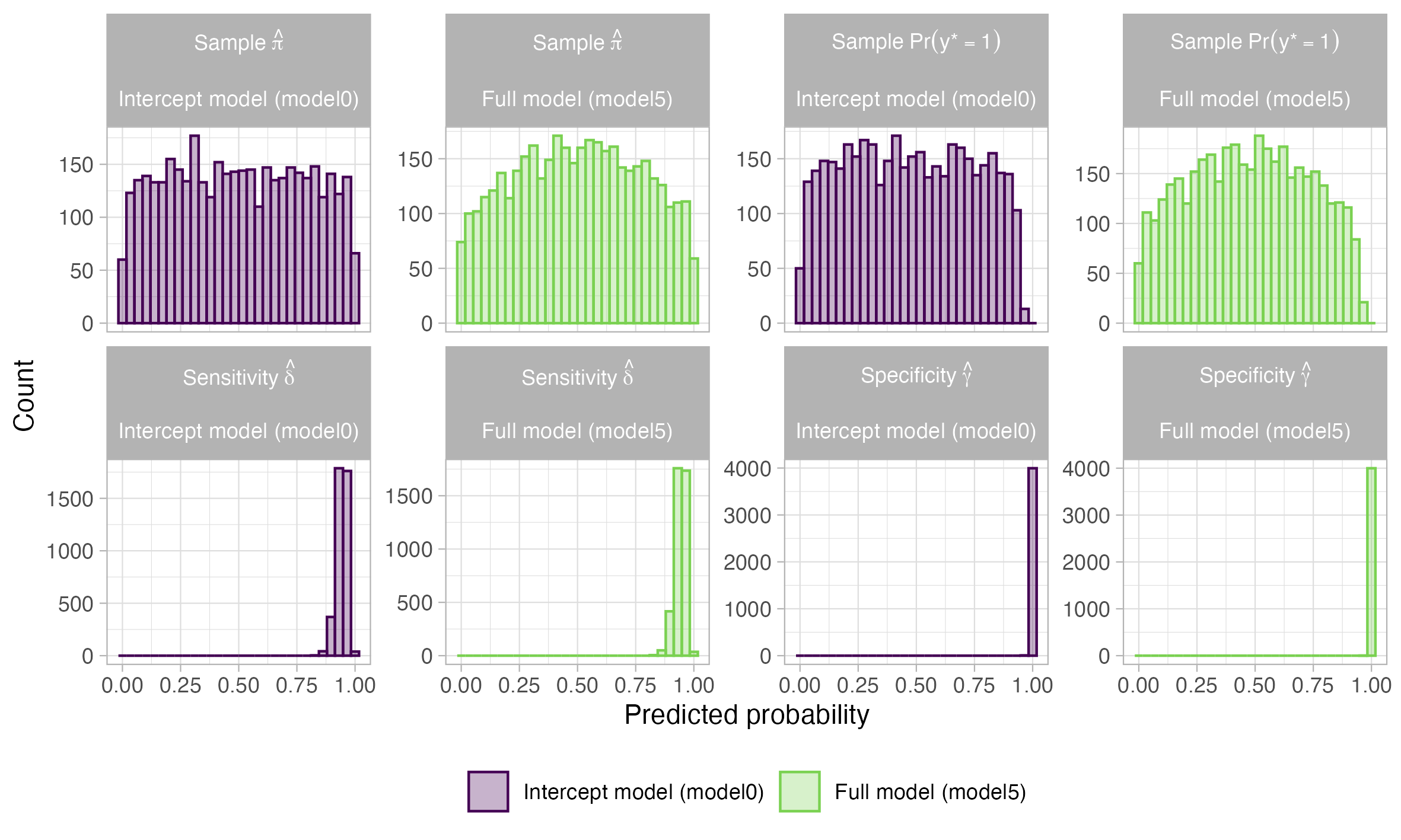}
\caption{\em Prior predictive distribution for overall population estimate of COVID-19 prevalence and observed test prevalence for the simplest (intercept-only) and more complex (with 5 covariates) model.}
\label{fig:real_example_prior_predictive}
\end{figure}

In Figure~\ref{fig:real_example_prior_predictive}, we see the prior predictive distributions for the population prevalence $\pi$, the sample prevalence $\mbox{Pr}(y^*=1)$, the sensitivity $\delta$, and the specificity $\gamma$. Here we see that as the model grows more complex (from model0 to model5), the prior predictive mass moves from a roughly uniform distribution across the probability space to a distribution that favors probability estimates in the center of the probability space more than the extreme ends near 0 and 1. 

If we observe a lower sample test prevalence than expected based on the prior, this could be the result of two aspects in the model. The first is our estimate of the overall test prevalence, $\widehat{\mbox{Pr}}(y^* = 1)$. While we see this to be true in model 0, in that our estimated test prevalence posterior has decreased from the prior, we do not see much impact in the posterior for test prevalence between model 0 and model 5, even though the prior predictive for these models changes considerably; see Figure~\ref{fig:real_example_prior_predictive}. 

The other parameter is much less intuitive. Consider an example with $100$ positive tests out of $200$. This is composed of the number of false positives and the number of true positives (positive tests = FP + TP).  Let's assume that sensitivity is $1$, so all positive cases are observed (no false negatives). In Table~\ref{tab:feedback_example}, we show two examples where the observed proportion of positive tests remains constant, but a changing specificity implies this is composed of a different number of false positives, which implies a different number of true positives and thus a different true prevalence. In Figure~\ref{fig:real_example_sens_spec_parameters}, we observe exactly this. The estimated observed tests remain constant over the models, but the specificity estimate decreases, implying a different true COVID-19 antibody prevalence.  

\begin{table}[]
    \begin{tabular}{cccccc}
       Sample size & +ve tests & Specificity & False positive & True positive & True prevalence\\
       \hline
      200 & 100 & 0.952 & 5 & 95 & 95/200 = 0.475 \\
      200 & 100 & 0.909 & 10 & 90 & 90/200 = 0.450
    \end{tabular}
    \caption{Simple example demonstrating how, if the number of tests remains constant but the specificity is changed, this implies a different true COVID-19 antibody prevalence.}
    \label{tab:feedback_example}
\end{table}

\section{Discussion}
In this work, we present a detailed and careful exploration of a model that incorporates measurement error within an MRP framework for the estimation of COVID-19 antibody prevalence. We demonstrate the use of a series of simulation studies to decompose and understand the contributing factors to unexpected results in a complex model. Ultimately, we cannot conclude a single source for unusual findings, but we believe the findings together are useful for those working with similar and adjacent models. 

We begin by demonstrating that the posterior predictive checks can be completed in an MRP setting. By specifically completing a posterior predictive check at the aggregate level, we avoid the challenges identified by Kuh et al.,\citep{kuh2024using} who demonstrate that individual predictive goodness does not necessarily align with predictive goodness at a population average level. By comparing the aggregate posterior predictive check to the sample mean, we can identify whether we specifically have a modeling issue or a poststratification issue. In this case, as the challenge was observed in the aggregate sample posterior predictive check, we concluded that we have a model rather than a poststratification issue. 

We identified three potential causes of the worsening posterior predictive check. We first explore the estimation of relatively rare prevalence with MRP. Rare-events bias is a demonstrated issue for logistic regressions,\citep{king2001logistic, leitgob2013, williams2016} but to our knowledge it has not been explored in an MRP context. Our first experiment showed that whilst the estimated intercept term is biased (consistent with existing literature) when prevalence is low and sample size is small, this does not reflect a bias in the estimation of the sample mean. This is consistent with other counterintuitive findings in MRP validation research.\citep{kuh2024using}

We then incorporated uncertainty in the tests when compared to the true COVID-19 antibody status. In the estimation of test error, we introduced measurement error estimation---particularly specificity---into our MRP model. When specificity is assumed to be known rather than estimated from data, the population prevalence estimate depends on the specificity value but is not affected by model complexity. When we add in a component of estimation for specificity, we then see that as the model grows more complex, the overall prevalence of the estimate decreases. This pattern is also observed in the real data,  where the degree of bias varies with the specificity value and the amount of available information ($m_\gamma$) for estimating specificity. 

In the third simulation, we incorporated additional predictors in model 3 onward, mirroring the later models in the original analysis. When little information was available on the test sensitivity and specificity (in other words, small $m_\gamma$), bias increased as more covariates were added to the hierarchical model. However, as $m_\gamma$ increases, the presence of one or two additional predictors no longer influences bias, and the increasing bias effect disappears. Similarly, in the real data, bias was less pronounced when non-varying effect predictors were not added to the model.

Our simulation studies reveal that when sensitivity and specificity are incorporated into an extended MRP model, the overall model prior (the overall suggested prevalence of the model) can feedback into the estimated test specificity, even if the data used to inform the test specificity directly do not change. This then causes a complex feedback loop that results in a decreased estimate for population prevalence as the model grows more complicated. 

Overall, our simulations suggest that a general deviance of the estimate of population prevalence when compared to the sample mean does not necessarily indicate poor model performance. In our simulations, we saw good recovery of population prevalence, and the deviation from the sample mean was an expected correction for measurement error. However, when specificity is estimated, we see the estimate for specificity decreases as the model grows more complex, suggesting feedback. We recommend that practitioners using a complex MRP model, such as the one presented in this paper, should examine layer by layer to check that results align with expectations and to disentangle any discrepancies that arise. As a starting point, comparisons against the sample average and poststratification on the sample data can be informative but should not be a gold standard when a measurement error model is used. When necessary, simulation studies should be conducted to validate the model’s behavior. 

In this work, we have aimed to disentangle the underlying causes of this bias and present a statistical workflow for diagnosing unexpected results in complex models. One potential solution to this problem is further work exploring global and regularizing priors to avoid increased feedback as the model grows more complex. In general, it is important to evaluate the statistical properties of adjustment procedures as the number of covariates increases. Our findings suggest a workflow for identifying and addressing this sort of systematic error in real-world analyses.

\section*{Acknowledgments}
We thank the U.S. National Institutes of Health and Office of Naval Research for partial support of this work, and Marnie Downes and John Carlin for providing the research problem and dataset, as well as for their insightful comments on early versions of this manuscript. 

Code to replicate the analyses and simulations presented in this paper are available here \url{https://github.com/swenk238/covidMRP_public}

\bibliographystyle{abbrvnat}
\bibliography{biblio}

\newpage 
\appendix

\section{Additional proofs and results}

\subsection[Solving for the intercept]{Solving for $\zeta_0$}\label{ssec:solveb0}

In equation (\ref{eq:pi_y_gen}), let $\pi_i = \mbox{Pr}(y_i = 1)$, taking logits on both sides and finding the expectation, we get:
\begin{eqnarray*}
    \mbox{E}(\logit(\pi)) &=& \mbox{E}(\zeta_0 + \zeta_1X_1 + \zeta_1X_2 + \dots + \zeta_1X_5) \\
 \mbox{E}(\logit(\pi)) &=& \mbox{E}(\zeta_0) + \mbox{E}(\zeta_1X_1) + \mbox{E}(\zeta_1X_2) + \dots + \mbox{E}(\zeta_1X_5)
\end{eqnarray*}
Since we set the values of $\zeta_1$ and generate $X$, we take the expectation of $X$ (which is not the case usually) to get
$$  \mbox{E}(\logit(\pi)) = \mbox{E}(\zeta_0) + \zeta_1\mbox{E}(X_1) + \zeta_1\mbox{E}(X_2) + \dots + \zeta_1\mbox{E}(X_5).$$
Since we are setting $\pi_i^* = \mbox{Pr}^*(y_i = 1)$ to be 0.01, $\zeta_1 = 0.3$ and generating five $X$'s from a uniform distribution between $-0.5$ and 0.5 (expected value of $\mbox{uniform}(a,b)$ is $(a+b)/2$),
\begin{eqnarray*}
-4.595 &=& \mbox{E}(\zeta_0) + 0.3 ( (-0.5+0.5) / 2) + 0.3 ( (-0.5+0.5    ) / 2)  + \dots + 0.3 ( (-0.5+0.5) / 2) \\
\mbox{E}(\zeta_0) &=& -4.59512 - 0.3 \cdot 0 - 0.3 \cdot 0 \dots - 0.3\cdot 0\\
\mbox{E}(\zeta_0) &=& -4.595.
\end{eqnarray*}
For a generic formula to find the expectation of $\zeta_0$ in our case, for $K$ is the number of covariates X:
$$ \mbox{E}(\zeta_0)  = \mbox{E}(\logit(\pi) + \zeta_1 * \sum_k(\mbox{E}(X))_{[k])},$$
But since we generate $X$ to have an expectation of 0,
$$ \mbox{E}(\zeta_0)  = \mbox{E}(\logit(\pi)). $$
To check that our calculation is correct, using the expected value of $\zeta_0$, we simulated to check that our expected value of $\pi^* = \mbox{Pr}^*(y_i = 1)$ is about 0.01.

\newpage 

\subsection[Proof of constraints]{Proof of constraints for $\mbox{Pr}(y^*_i = 1)$}\label{ssec:proof of constraints}

Let $\delta$ = sensitivity and $\gamma$ = specificity.  We denote $\mbox{Pr}(y_i^*=1)$ as $p$ and $\pi_i$ as $\pi$ for simplicity. Solving for $p$ by using $0 \leq \pi \leq 1$:
\begin{align*}
p &= \pi\delta + (1 - \pi)(1 - \gamma) \\
\pi &= \frac{(p + \gamma - 1)}{(\delta + \gamma - 1)}; \ \ 0 \leq \pi \leq 1 \\
0 &\leq \frac{(p + \gamma - 1)}{(\delta + \gamma - 1)} \leq 1 \\ 
\end{align*}
Since we need to multiply the denominator $(\delta + \gamma - 1)$, we investigate three scenarios:
(a) When $\delta + \gamma - 1 > 0$ or $\delta > 1 - \gamma$: \\
\begin{align*}
    0 &\leq (p + \gamma - 1) \leq (\delta + \gamma - 1) \\ 
    1 &\leq p + \gamma \leq \delta + \gamma \\
    (1 - \gamma) &\leq p \leq \delta  
    \\
\end{align*}
(b) When $\delta + \gamma - 1 < 0$ or $\delta < 1 - \gamma$: 
\begin{align*}
    (\delta + \gamma - 1) &\leqslant (p + \gamma - 1) \leqslant 0 \\ 
     \delta + \gamma &\leqslant p + \gamma \leqslant 1\\
    \delta &\leq p \leq (1 - \gamma)     
\end{align*}
(c) When $\delta = (1 - \gamma)$, the denominator becomes 0 and the fraction becomes undefined.

Thus when $\pi_i = \mbox{Pr}(y_i=1)$---the probability of the true COVID-19 antibody status of an individual is positive---is constrained within $(0,1)$, then $\hat{p}$, or $\widehat{\mbox{Pr}}(y^*_i = 1)$, the expected frequency of an individual testing positive must be constrained within 

\begin{equation*}
\begin{cases}
        (1 - \gamma)  \leq \widehat{\mbox{Pr}}(y_i^*=1) \leq \delta,& \text{if } \delta > (1-\gamma)\\
        \delta \leq \widehat{\mbox{Pr}}(y_i^*=1) \leq (1 - \gamma) ,& \text{if } \delta < (1-\gamma) \\
        \infty,              & \text{if } \delta = (1-\gamma).
 \end{cases}
\end{equation*}

\newpage 

\subsection{Modifications to the real data model in different experimental settings}

\renewcommand{\arraystretch}{2}
\begin{table}[!htb]
\begin{threeparttable}
\resizebox{1.05\textwidth}{!}{
\begin{tabular}{p{0.1cm}>{\centering\arraybackslash}p{0.1cm}c>{\centering\arraybackslash}p{0.1cm}>{\centering\arraybackslash}p{2cm}p{2.3cm}p{2.5cm}p{2.6cm}p{2.8cm}}
& & & \multicolumn{2}{p{3.5cm}}{\bfseries Experiment I: rare-events bias in logistic models (with covariates)} 
& \multicolumn{2}{p{4.8cm}}{\bfseries Experiment II: Inclusion of the uncertainty of test sensitivity and specificity} 
& \multicolumn{2}{p{5.4cm}}{\bfseries Experiment III: Adding overall effects terms} \\
\renewcommand{\arraystretch}{3}
& & \multirow{3}{*}{\bfseries Model} & \multicolumn{1}{l}{\textbf{Exp 1.1}} &  \textbf{Exp 1.2} &  \textbf{Exp 2.1} &  \textbf{Exp 2.2} &  \multicolumn{2}{c}{\textbf{Exp 3}} \\
& & & \multirow{3}{=}{\textbf{intercept-only models}} & \multirow{3}{=}{\textbf{Without overall effects and  measurement error\tablefootnote{Equivalent to a classic multilevel logistic regression using equations (4)--(5) only.}}} & \multirow{3}{=}{\textbf{Without overall effects but known specificity}} & \multirow{3}{=}{\textbf{Without overall effects but estimating specificity\tablefootnote{Different sample sizes for true positives, true negatives, false positives and false negatives.}}} & \multicolumn{2}{c}{{\textbf{Removing overall effects terms}}}  \\
& & & & & & &  \textbf{No overall effects term}& \textbf{One overall effects terms}\\ \\
\hline
\multirow{7}{*}{\rotatebox{90}{\textbf{Covariates added sequentially}}} & \multirow{8}{*}{\tikz \draw[<-, thick] (0,15)--(0,22);} & \textbf{0} & - & $\pi_i = \mbox{logit}^{-1}\big(\beta_0\big)$ & \multirow{6}{2.5cm}{\vfil In equation (6), fixed sensitivity value at 1, and specificity at fixed values = $\{0.98$, $0.990$, $0.995$, $1.000\}$ respectively.} & \multirow{6}{2.5cm}{\vfil Two approaches: (1) Keep sensitivity fixed at 1 while estimation of specificity with a beta prior using varying sample sizes, (2) estimate both specificity and sensitivity with beta priors and varying sample sizes.} & No change & No change \\ 
& & \textbf{1} & - & $\pi_i = \mbox{logit}^{-1}\big(\beta_0 + \alpha_{\text{strata}}\big)$ & & & No change & No change \\
& & \textbf{2} & - & $\ldots + \alpha_{\text{postcode}}$  & & & No change & No change \\ 
& & \textbf{3} & - &$  \ldots + \alpha_{\text{seifa}} $  & & & \multirow{3}{=}{Continuous SEIFA variable removed} & No change \\    
& & \textbf{4} & - &  No change\tablefootnote{Binary sex variable was added at this stage in the original application. Since overall effects are not added, model 4 remains the same as model 3 in this experiment setup.}  & & &  & \multirow{2}{=}{Binary sex variable removed} \\ 
& & \textbf{5} & - & $\ldots + \alpha_{\text{age}}$  & &  & & \\
&  \multicolumn{8}{r}{\tikz \draw[<-, thick] (2,1) -- (17,1);\vspace{-5mm}} \\
\multicolumn{9}{c}{\textbf{Variables are gradually removed}} 

\end{tabular}
}
\caption{\em In each experiment, parts of the model (or covariates) are gradually removed from Models 0 through 5 to mimic the settings in the simulation setup.}
\label{tab:models_fit_realdata}
\end{threeparttable}
\end{table}

\newpage 

\subsection{All iteration results from Experiment 1.1: Intercept-only model} \label{ssec:alliter}
\begin{figure}[!htb]
\centering
\includegraphics[width=0.95\textwidth]{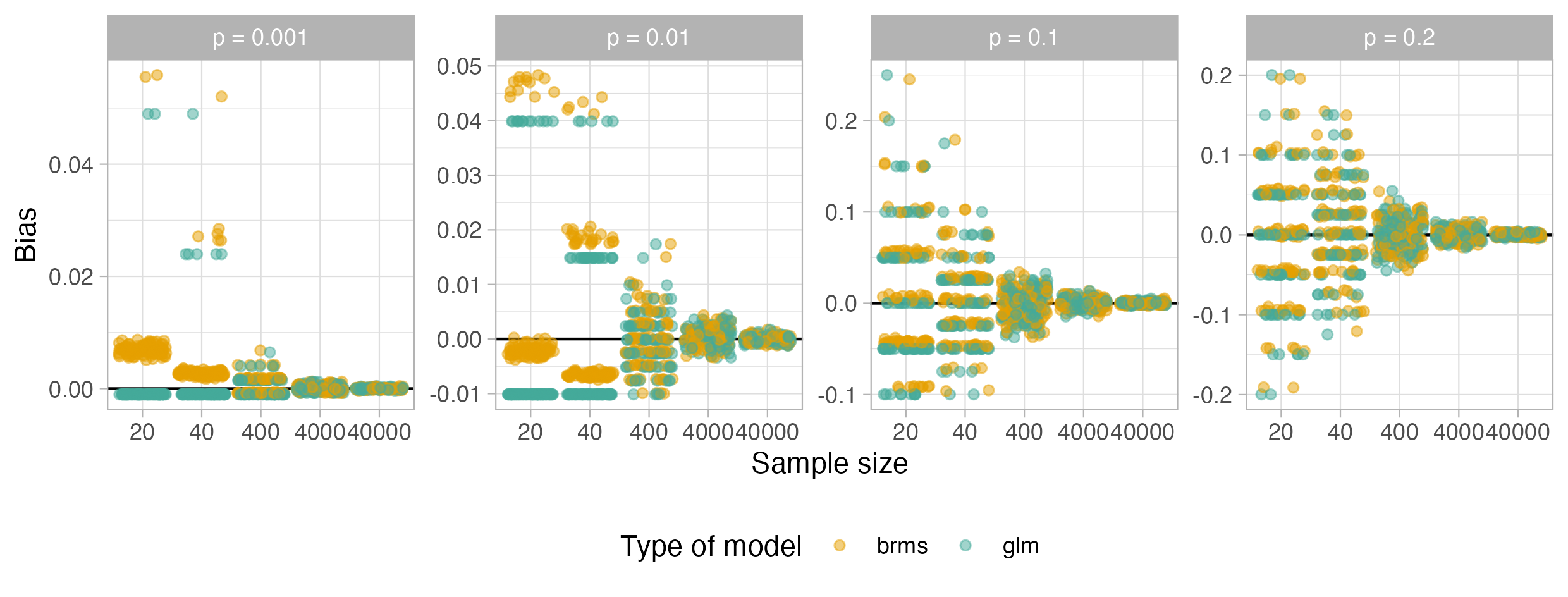}\vspace{-1.5em}
\caption{\em Bias ($y$-axis) for predicted estimates using varying sample sizes ($x$-axis) from the intercept-only models, for true probability of outcome $= 0.001, 0.01, 0.1, 0.2$ in each panel. Each point represents a single iteration. Unlike the box plots presented in the main, which show overall expected bias, this figure shows the clumping of model estimates with smaller sample sizes. This is driven by the properties of the sample and the number of observations that were 1 in that particular iteration. Not that most cases when sample size is small and the prevalence is rare are in samples where no positive observations were observed, with the ``outliers'' where at least 1 observed. }
\label{fig:brms}
\end{figure}

\newpage
\subsection[With covariates]{Experiment 1.2: With covariates: $\zeta_1 = 0.3$}\label{ssec:covbeta0.3}

\begin{figure}[!htb]
\centering
\includegraphics[width=0.7\textwidth]{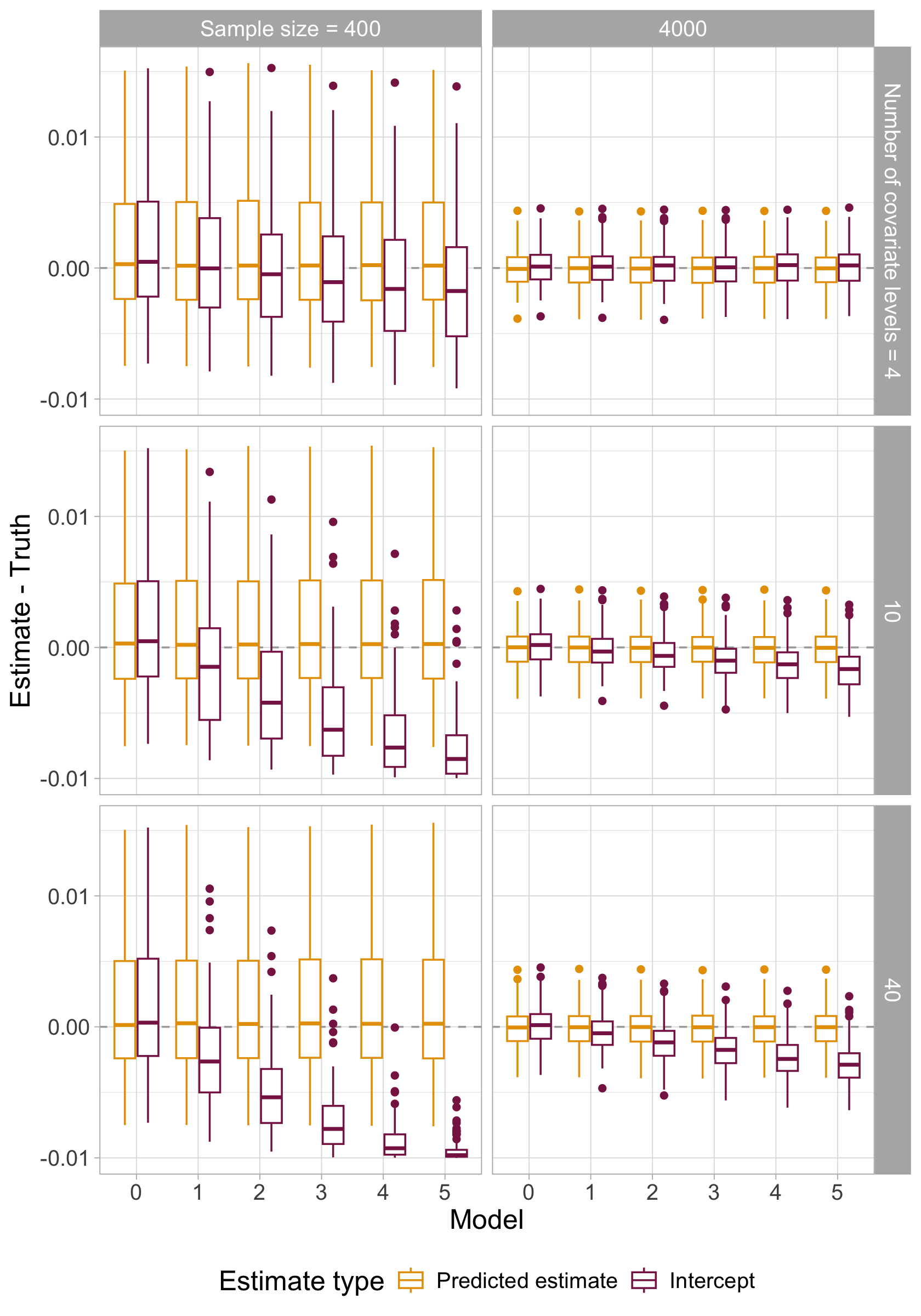}
\caption{\em Box plots for bias ($y$-axis) for predicted estimates and intercept using models with sequentially added covariates ($x$-axis), for sample size of 400 (left panel) and 4000 (right panel) and each with a different number of covariates  (4, 10, or 40) on the horizontal panel, fixing $\zeta_1 = 0.3$. All the models are fitted using the \texttt{brm} function, which calls Stan from R. Orange is for the predicted estimate, and purple is for the predicted intercept. The plot shows intercept gets increasingly biased when we are estimating more number of levels in each covariate. The pattern is more apparent when the sample size is 400.}
\end{figure}

\newpage
\subsection[With covariates]{Experiment 1.2: With covariates: $\zeta_1 = 0$}\label{ssec:covbeta0}
\begin{figure}[!htb]
\centering
\includegraphics[width=0.7\textwidth]{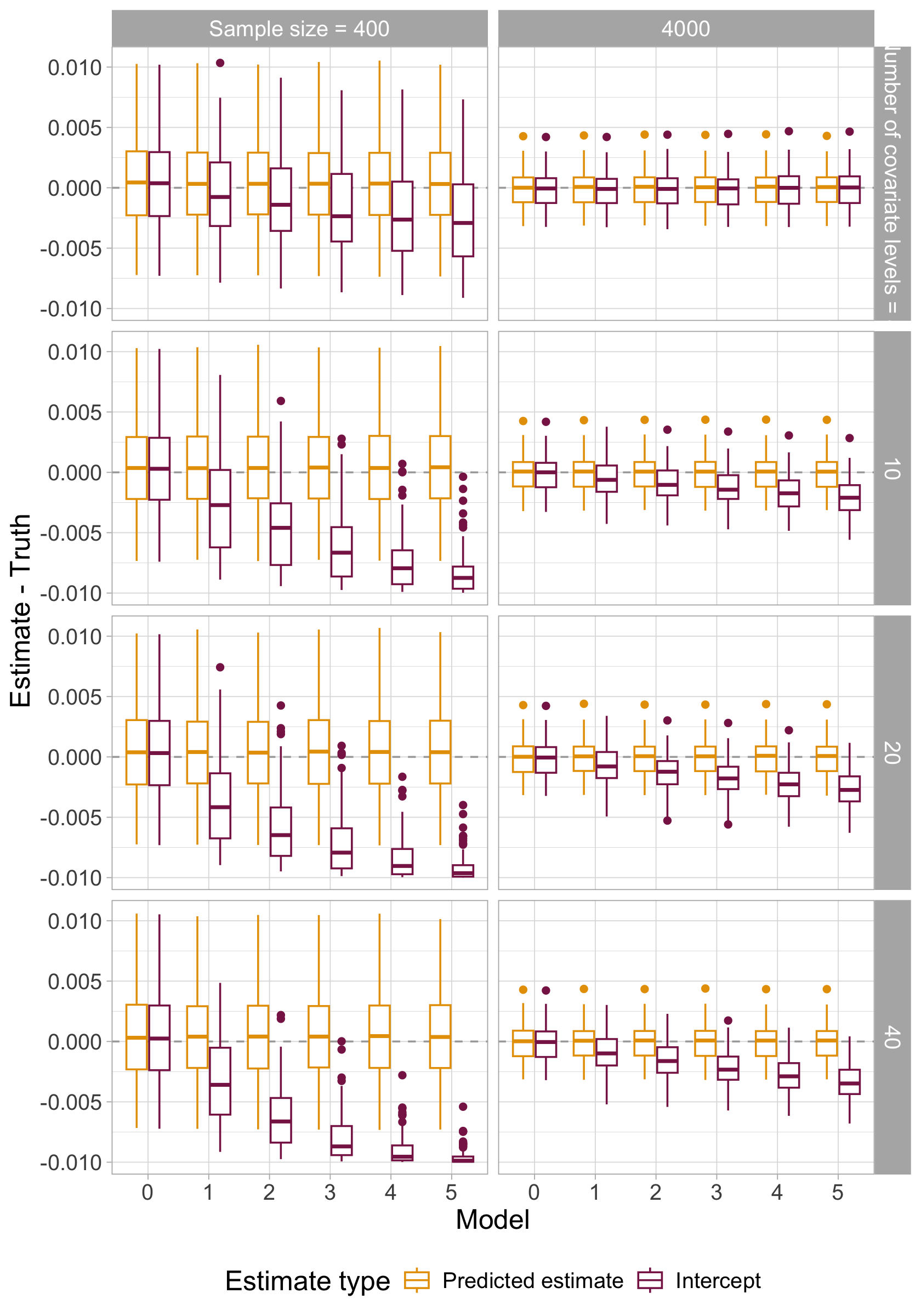}
\caption{\em Box plots illustrating bias for predicted estimates and intercept ($y$-axis) using models with sequentially added covariates ($x$-axis), for sample size $= 400$ (left panel) and 4000 (right panel). Each horizontal panel corresponds to a different number of covariate levels (4, 10, or 40) with $\zeta_1$ set to 0. All models are fitted using the the \texttt{brm} function, which calls Stan from R. Predicted estimates are shown in orange, and intercept estimates are shown in purple. The plot indicates that intercept bias increases as the number of covariate levels increases, with this pattern being more pronounced when the sample size is 400.}
\end{figure}

\FloatBarrier

\newpage
\subsection{Experiment 2.1: other covariate levels when not estimating specificity}\label{ssec:noEstSpec_app}
\begin{figure}[!htb]
\centering
\includegraphics[width=0.9\textwidth]{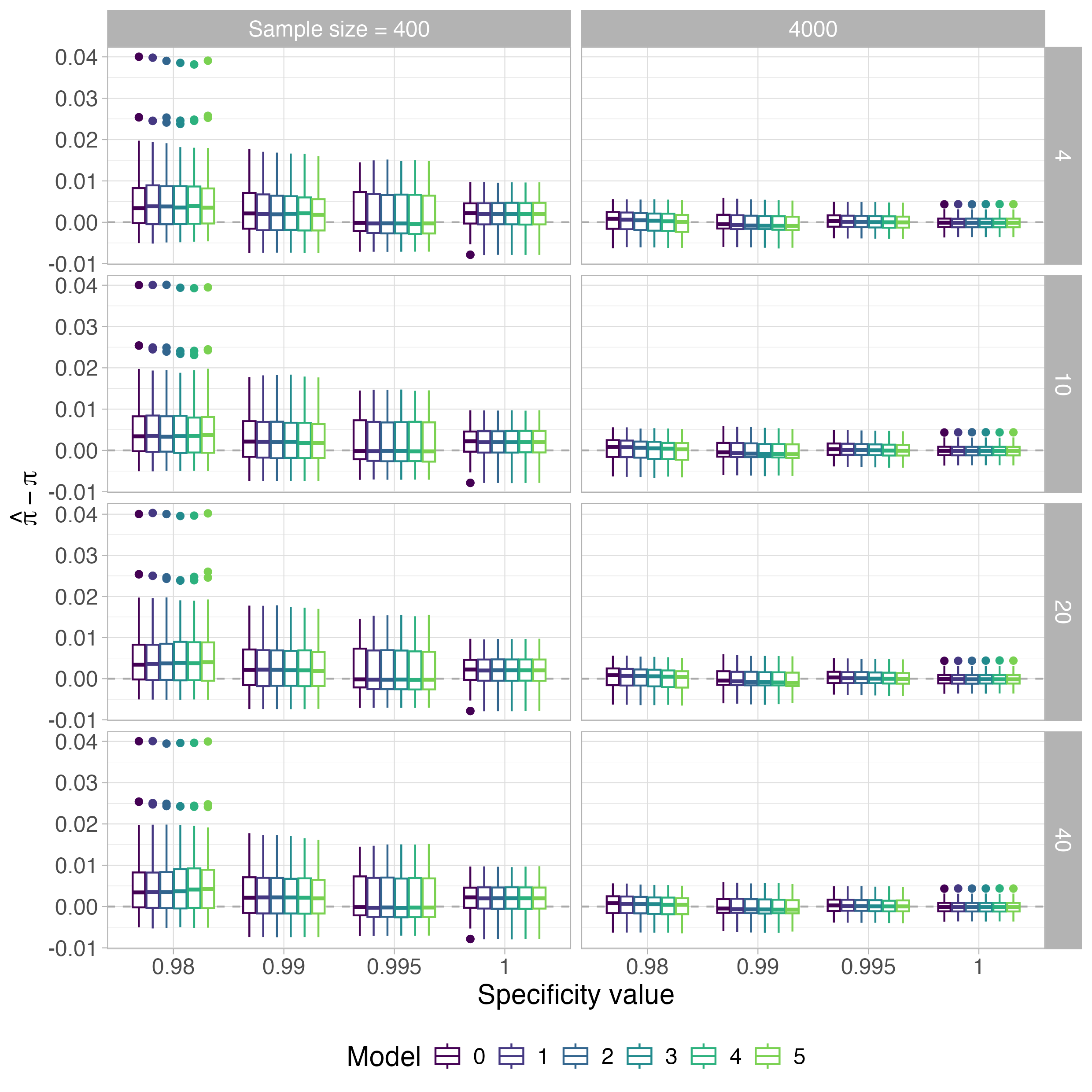}\vspace{-2em}
\caption{\em Difference between estimated COVID-19 antibody prevalence and true COVID-19 antibody prevalence ($y$-axis) using models with sequentially added covariates (color, lighter shade indicating more complex model), for sample size $= 400$ (left panel) and 4000 (right panel) and each with a different number of covariate levels (4, 10, or 40) on the horizontal panel, fixing $\zeta_1 = 0.3$. The different specificity values are included on the $x$-axis. The plot shows that when estimating the true COVID-19 antibody prevalence the model is relatively unbiased, although with some overestimation with smaller sample sizes. The number of levels in each covariate does not seem to have made an impact.}
\end{figure}

\newpage

\begin{figure}[!htb]
\centering
\includegraphics[width=0.9\textwidth]{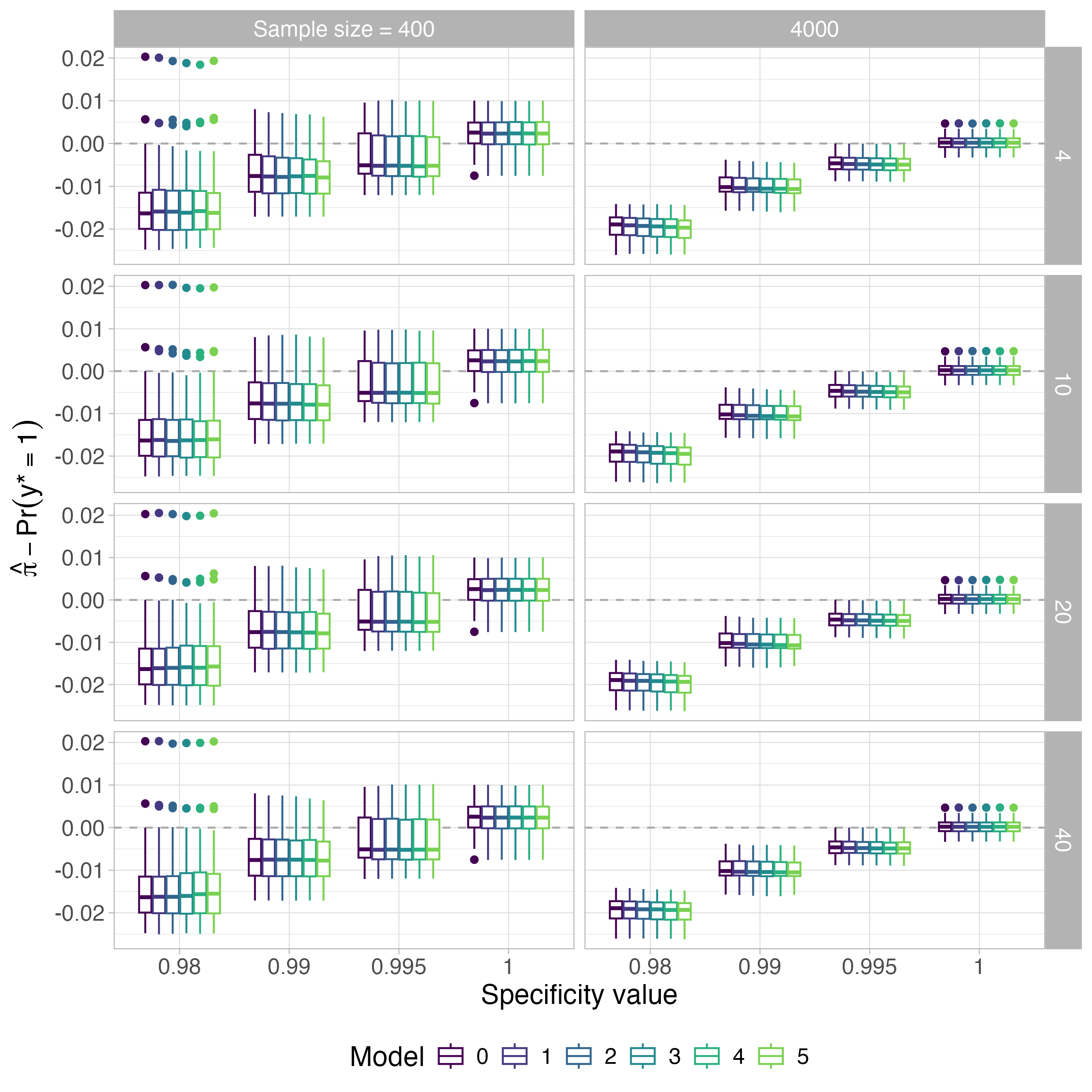}\vspace{-2em}
\caption{\em Difference between estimated COVID-19 antibody prevalence and true test prevalence using models with sequentially added covariates (color, lighter shade indicating more complex model), for sample size $= 400$ (left panel) and 4000 (right panel) and each with a different number of covariate levels (4, 10, or 40) on the horizontal panel, fixing $\zeta_1 = 0.3$. The different specificity values are included on the $x$-axis. The plot shows that if we compare the estimate COVID-19 antibody prevalence to the true test prevalence, we see a consistent difference that depend on specificity. The number of levels in each covariate does not seem to have made an impact.}
\end{figure}

\newpage

\subsection{Experiment 2.2: other covariate levels when estimating specificity}

\begin{figure}[!htb]
\centering
\includegraphics[width=0.9\textwidth]{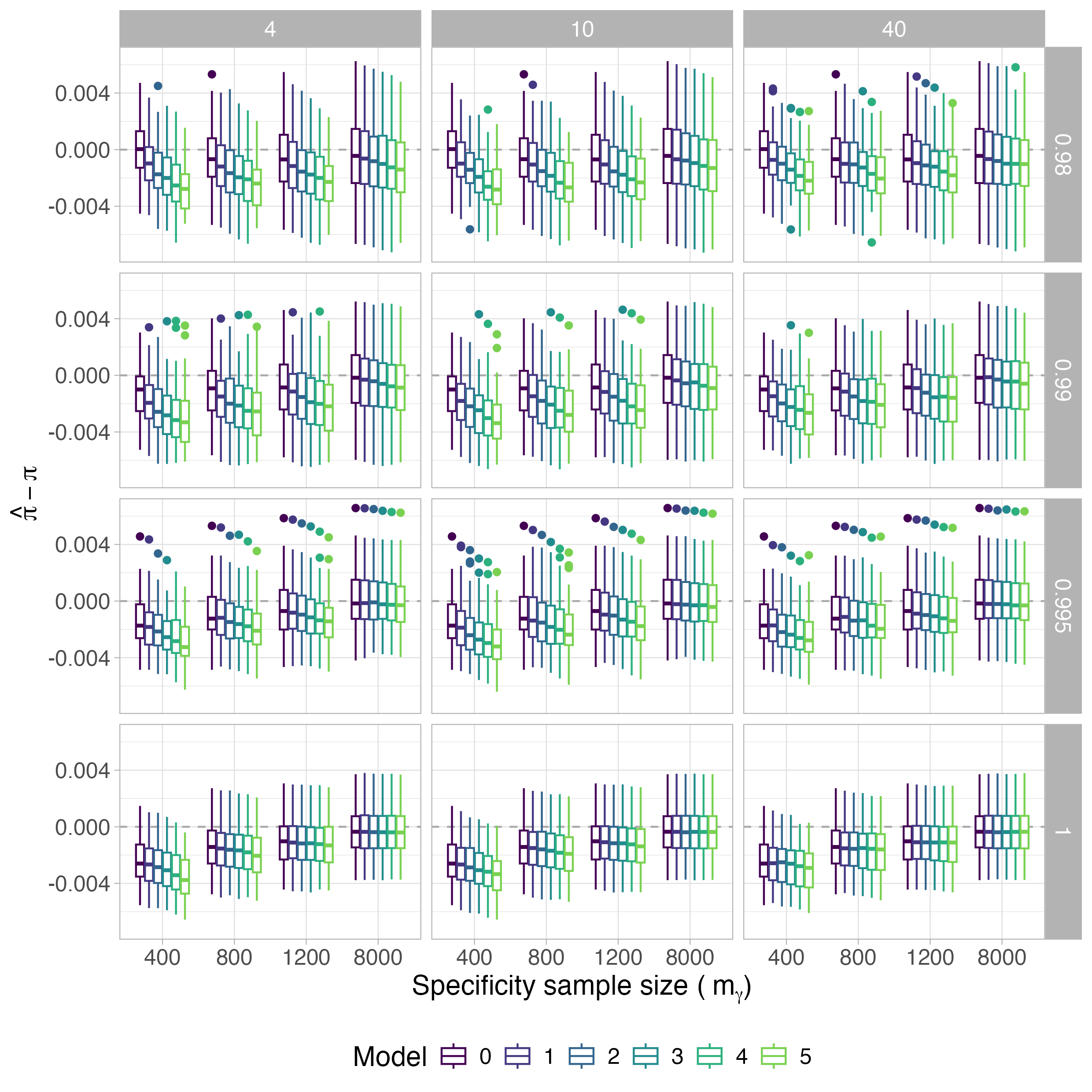}
\caption{\em Bias ($y$-axis) for predicted estimates using models with sequentially added covariates (represented by color), for various specificity values on the horizontal panel and a different number of covariate levels (4, 10, or 40) on the vertical panel, fixing $\zeta_1 = 0.3$. The vertical panels represent different specificity values, while the $x$-axis represents the sample size $m_\gamma$ used to estimate specificity. All the models are fitted using
customized Stan code. While the true specificity and data used to estimate the true specificity both impact either the change in bias as models grow more complex (sample size) or the initial bias (specificity), the number of categories in each level isn't as impactful.}
\end{figure}
\newpage

\subsection{Experiment 3. Impact of sample size used to estimate specificity. }\label{ssec:exp3_app}

\begin{figure}[!htb]\centering
    \includegraphics[width=0.9\textwidth]{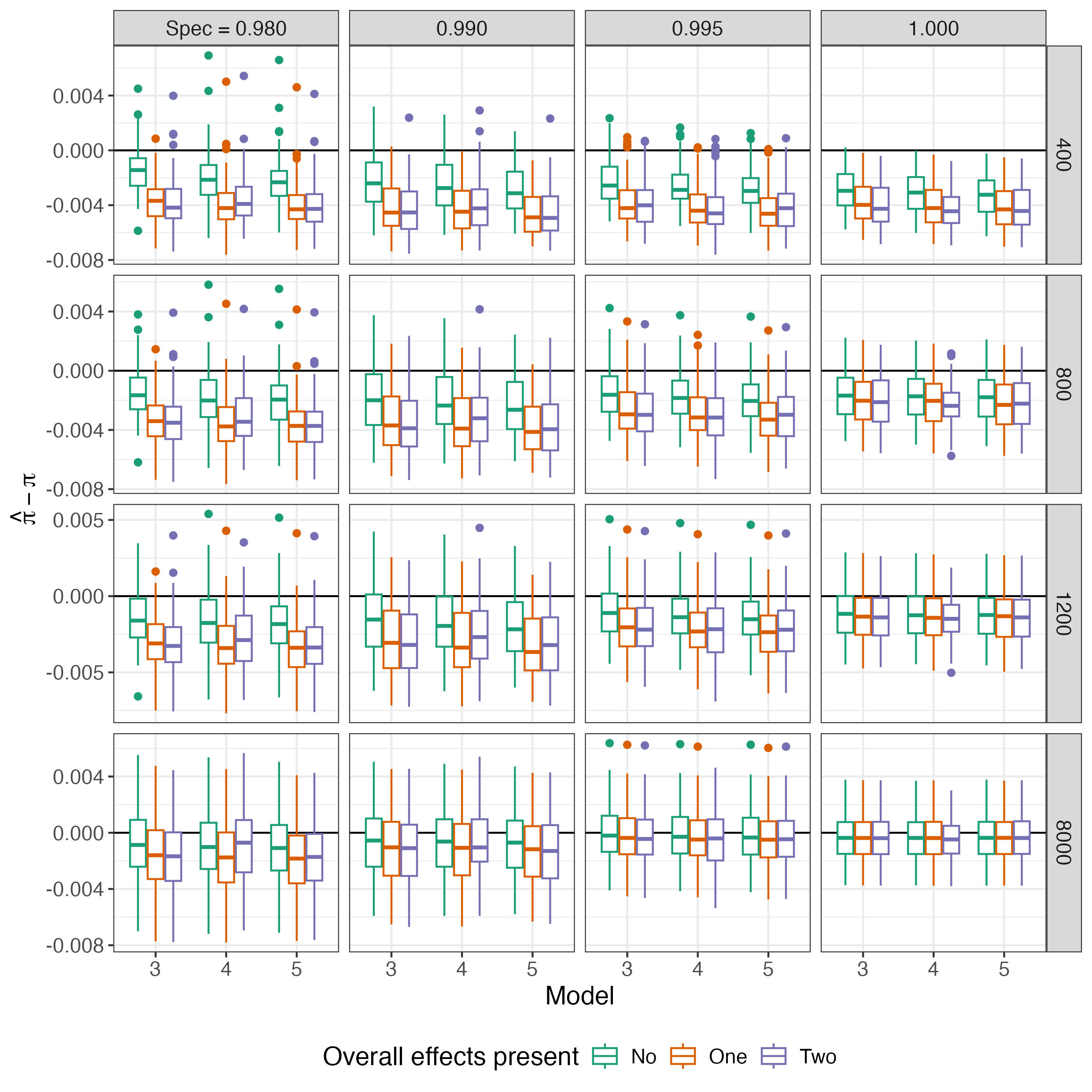} 
     \setlength{\abovecaptionskip}{-5pt}
    \caption{\em Difference in the bias of predicted estimates with model 0 ($y$-axis) using models with sequentially added covariates, with specificity estimated from external data and overall effects terms added ($x$-axis), for a sample size of 4000 and 20 covariate levels for 100 iterations. We add either one or two overall slope coefficients (fixed effects) into Model 3 onward . The vertical panels are when the specificity value changes, and the horizontal panels are for the various prior sample sizes $m_\gamma = 400, 800, 1200,$ and $8000$.}
    \label{fig:diff_m0_all}
\end{figure}

\end{document}